\documentclass[aps]{revtex4}
\usepackage{amssymb,amsmath,bm,graphicx,color,subfigure}
\begin{document}
\title{L\'{e}vy flights and nonlocal  quantum dynamics}
\author{Piotr Garbaczewski  and  Vladimir Stephanovich}
\affiliation{Institute of Physics, University of Opole, 45-052 Opole, Poland}
\date{\today }
\begin{abstract}
We develop a fully fledged theory of   quantum dynamical patterns of behavior that are
 nonlocally induced.  To this end we generalize the standard
 Laplacian-based framework of the Schr\"{o}dinger picture quantum  evolution
  to that employing   nonlocal (pseudodifferential)  operators.
  Special attention  is paid to the Salpeter (here, $m\geq 0$)  quasirelativistic equation and the
    evolution of  various  wave packets, in particular to their  radial expansion in 3D.
  Foldy's synthesis of "covariant  particle    equations" is extended to
      encompass free  Maxwell theory,  which  however  is devoid of any "particle" content.
  Links with  the  photon wave mechanics are explored.
        \end{abstract}
\maketitle
\section{Introduction}

The  standard unitary  quantum  dynamics   and the  Schr\"{o}dinger semigroup-driven   random   motion, \cite{risken,albeverio}, are examples
of dual  evolution scenarios that may be mapped among each other  by means of a suitable   analytic continuation in time
procedure.  This is an offspring   \cite{gar0,guerra,guerra1}  of   Euclidean quantum field theory  methods,
 albeit  reduced to  the  purely  quantum  mechanical level.
  Both  evolutions  are generated  by  means of  a  common   local Hamiltonian  operator.
 Our departure point for subsequent analysis is an observation that a  complete   spectral resolution
   of the corresponding    Hamiltonian   actually  determines a   classical  space-time homogeneous  diffusion-type
    Markov  process  in $R^n$.

Within the general  theory of   so called infnitely divisible probability laws (see below)    the familiar
 Laplacian (Wiener noise or Brownian motion generator)  is known to be one isolated member  of a  surprisingly
  rich family of non-Gaussian L\'{e}vy  noise generators.   All of them
stem  from the  fundamental  L\'{e}vy-Khintchine formula,  and typically refer to
  probability distributions of spatial  jumps and the resultant jump-type Markov processes.
   That needs to be contrasted with the traditional diffusion imagery    (Wiener noise and process)
   associated with the Laplacian,  \cite{nelson}.

 The emergent  L\'{e}vy generators  are  manifestly  nonlocal (pseudo-differential) operators and, while being  additively perturbed by
 a suitable external potential,   give rise   (via a canonical quantization procedure described subsequently)
   to  L\'{e}vy-Schr\"{o}dinger semigroups.
  The  dual image of such  semigroups   comprises  unitary dynamics scenarios  which  can be viewed
   as   signatures of  a nonlocal    quantum behavior.  As we discuss in below this dynamical
    nonlocality extends to the very    concept of photons, within  so-called  photon wave mechanics, \cite{ibb}.

Quite apart from  "Euclidean" vs "real" time   connotations,  the considered   dual dynamical systems
  refer to   real time labels and clocks.  Both of them drive   probability density functions (pdfs) which,
in  the course of time evolution ($t\in R_+$)     either maintain or    develop  asymptotic   heavy-tails and typically have
 a finite number of moments in existence.  This  needs to be contrasted with the Gaussian standards
 of thinking  (all  pdf moments in existence, rapid decay at infinities etc.)  that pervade the   Laplacian-based  quantum theory.

 The major  goal of  the present paper is to set on solid grounds the quantization programme  that  completely avoids a
 reference to classical mechanics  (of massive particles),  normally  viewed  as a conceptual   support for  the specific choice of
 the Hamiltonian operator  within the   traditional   Schr\"{o}dinger wave mechanics.
Here we consider a  standard  form of the Hamiltonian (minus Laplacian plus a perturbing potential)  as an
exception  rather  than  a   universally  valid  feature  of  an admissible   quantum theory.

 The latter is introduced by  means of the  most
primitive  quantization ansatz,   whose core lies in choosing  the Hilbert space $L^2(R^n)$ as an arena
 for our investigations. From the start we have the Fourier transformation realized as a unitary operation
  in this space and a canonical quantization input as a straightforward consequence.

The above mentioned   L\'{e}vy-Khintchine formula, while  being  tailored to our purposes, derives through a
Fourier transform of a symmetric probability density function.
A  variety of symmetric  probability laws for  random noise  is classified by means of a characteristic function
 which is an  exponent  $\eta (p)$     of the $(2\pi )^{n/2}$-multiplied  Fourier transform of that pdf:
$\int d^nx \rho (x) \exp(\pm ipx) = \exp [\eta (p)]$.

The  most straightforward,  naive canonical  quantization step
   $p\rightarrow   \hat{p} =-i\nabla $   (hereby limited to  the  symmetric pdf   case)   introduces the notion of  random
  transport  that is   driven by L\'{e}vy-Schr\"{o}dinger semigroups  $\exp [ t \eta (\hat{p})]$.  Their dual partners actually   are  the
unitary evolution   operators $\exp [i t \eta (\hat{p})]$  of interest and are a subject of further discussion.

In the present paper we develop  a  complete theory  that starts  from  diffusion type processes and their semigroup reconstruction,
followed by the duality mapping.
Next we pass to  general   jump-type stochastic processes  that  stem from the L\'{e}vy-Khintchine formula and  pass to the dual (e.g. quantum)
 dynamics, with  a number of specific  physically motivated examples, like the Salpeter Hamiltonian and its $m=0$ Cauchy
  version. The major  departure point for the present
 analysis is Ref.  \cite{gar}, but  earlier references of relevance can be traced back therein, see also \cite{angelis}.

Our main objective is  to deduce and next expose   a fully nonlocal pattern of dynamical  behavior
appropriate for the description of  general   quantum  dynamics.  We believe that it is nowadays necessary  to
     take under scrutiny customary ways of thinking about quantum phenomena and open  new  conceptual    avenues, based
on the inherent nonlocality   of the dynamics generator, c.f.   \cite{lammerzahl} for an earlier attempt.

\section{Diffusion-type  processes vs   quantum motion}
\subsection{Ground state condition}

Taking as obvious the standard wisdom  about the Schr\"{o}dinger picture quantum dynamics, we merely recall
(here, without  specifying  the space  dimension) that the equation
\begin{equation}
  i\hbar \partial _t \psi  = \left[ - {\frac{\hbar ^2}{2m}} \Delta +
V \right] \psi  = \hat{H} \psi  \label{schr}
\end{equation}
involves a Hamiltonian
$\hat{H}$  that is a self-adjoint  operator in a suitable Hilbert space domain; we
consider time-independent potential $V=V(x)$.
Since we are interested in non-negative  generators of motion,
for later convenience we  impose an additive renormalization of a bounded from below  Hamiltonian   (to be absorbed in the explicit
 functional expression for  $V$),  such  that
 $\hat{H} \geq 0$  and $0$ is its lowest eigenvalue.

  Further standard notations  are reproduced  for the record:  for any normalized $\psi$,     $\rho (x,t)= |\psi |^2(x,t)$ stands for a pdf,
$v= (\hbar /2mi)[ (\nabla \psi /\psi ) -  (\nabla \psi ^*/\psi ^*)]$ is   a current velocity field, while $j= v \rho $
 a probability current: $\partial _t\rho = - \nabla \, j$.
 While keeping  the polar (Madelung) decomposition of $\psi $ in mind , we consider $v$ in the gradient form: $v= {\frac{1}m} \nabla s$.

We note that by introducing  $u(x,t) \doteq (\hbar / 2m)\, \nabla \ln \rho   $  and next the drift field $ b(x,t) = v + u$, we can  rewrite the
continuity equation in the Fokker-Planck form  $ \partial _t \rho = D\, \Delta \rho  -\nabla  (b \rho )$,
 with $D={\frac{\hbar}{2m}}$  playing the role of the diffusion coefficient. Here, in general $b=b(x,t)$ is time-dependent, unless we  pass to stationary states.

  The ground state condition  for   $\hat{H}$  corresponding to
 the  lowest eigenvalue $0$,   i. e.   $\hat{H} \psi =0$  with $|\psi |= \rho_*^{1/2}$,
  directly involves the  negative  of the familiar de  Broglie-Bohm  "quantum potential"
\begin{equation}
 V= 2mD^2\,  {\frac{\Delta \rho _*^{1/2}}{\rho _*^{1/2}}}\, .
\end{equation}
Here $D=\hbar /2m$  is implicit, although not a must if out of the quantum context (like in the  case of  Brownian motion).

Having $\rho _*^{1/2}$, we have  completely   determined $V = V[\rho  _*]$, which  in turn  allows for  a  complete  reconstruction of
 the  (Schr\"{o}dinger)  semigroup  $\exp(-t\hat{H}/2mD)$, \cite{vilela1,vilela2}.
A traditional well established procedure   actually amounts to the reverse: given  the  (Schr\"{o}dinger)  semigroup with   a priori  chosen
 (admissible)  potential  $V(x)$, one    infers   $\rho _*^{1/2} [V] = \rho _*^{1/2} $ as the lowest strictly positive  eigenfunction of
 the   semigroup generator (i.e. of   $- \hat{H}/2mD$).

\subsection{Schr\"{o}dinger semigroup vs Fokker-Planck dynamics}

We keep the notation  $D$  for  a diffusion coefficient.
  The (Schr\"{o}dinger)    semigroup induces  the  related  generalized diffusion equation  (by setting  $V=0$ we get the
  standard heat equation)
 \begin{equation}
\exp(-t\hat{H}/2mD)\Psi _0 = \Psi  \, \,  \Longrightarrow   \, \,   \partial _t \Psi = \left[ D \Delta  -  {\frac{V}{2mD}}\right] \Psi  =
- {\frac{\hat {H}}{2mD}}\,  \Psi   \label{diff}
\end{equation}
This  Schr\"{o}dinger-type equation (albeit without an imaginary  unit $i$) employs $\hat{H}$ as  a  self-adjoint  relative of the
 Fokker-Planck operator,  \cite{risken},
\begin{equation}
  L_{FP}=  D\Delta  - \nabla (b\,  \cdot )\,     \Longrightarrow   \partial _t\rho = L_{FP}\, \rho
  \end{equation}
    and thence  indirectly determines the time
  evolution of   a   pdf   $\rho (x,t)$ for $t\geq 0$, the initial data being tacitly presumed.

To execute a transformation between the Fokker-Planck and  the  Schr\"{o}dinger  semigroup dynamics,
 let us  assume that the pertinent FP pdf   $\rho (x,t)$ has a strictly positive  asymptotics   $\rho _*(x)$.
We make an  ansatz
\begin{equation}
\Psi (x,t) = {\frac{\rho (x,t)}{\sqrt{\rho _*(x)}}} \, .
\end{equation}
  Additionally, we identify   $ b= D\nabla \ln \rho _* $ as a time-independent forward drift, and introduce
   (typically time-dependent)  osmotic   $ u= D\nabla \ln \rho $  and current   $ v= b- u = (1/m)\nabla s $  velocity fields.

 The connection between the Fokker-Planck and
 semigroup dynamics is readily   established, provided a  compatibility condition  Eq. (2) is extended to the drift field $ b= D\nabla \ln \rho _* $
\begin{equation}
 V(x) =  2mD^2\left[ {\frac{b^2}{2D}} + \nabla b\right]\, .
 \end{equation}
 It is clear that $\Psi (x,t)$ solves  the generalized diffusion equation  (3)   if the identity (5) is valid.
Given $V(x)$, we can in principle solve the Riccati-type equation  (6)  with respect to $b(x)$. Proceeding in reverse,
  with a pre-defined  forward drift $b(x)$ of a diffusion-type process, we have determined  an  admissible form of
  the Schr\"{o}dinger  semigroup potential $V(x)$.

We point out that in the  theory of diffusion processes, an impact of external  conservative  forces is  typically encoded in the drift
   function  $b \sim   -\nabla U$ where $U$ is a Newtonian potential. We emphasize  a substantial difference of functional
    forms for  $V$ and $U$ \, (except for the  notorious  harmonic oscillator  case).

The preceding  discussion  may be summarized by a   symbolic   identity,
 appropriate for operators acting in their  domains. Namely, we have  \cite{risken,vilela2}
\begin{equation}
-  \rho _*^{-1/2}\, L_{FP} \, \rho _*^{1/2} \equiv {\frac{1}{2mD}}\, \hat{H}\, . \label{mapping}
\end{equation}
where $\hat{H}\rho _*^{1/2} =0$.
On the other hand, if we consider  $\hat{H}$ as given a priori, then  the formula
  $L_{FP} = - \rho _*^{1/2} \hat{H} \rho _*^{-1/2}/ 2mD$ associates   a corresponding
 Fokker-Planck generator  $L_{FP}$   with  the Hamiltonian $\hat{H}$.

It is worthwhile to mention the  equivalence of  Eq. (\ref{mapping})  and the  transformation recipe:
\begin{equation}
\partial _t \rho =D\,  [\rho ^{1/2}_* \Delta (\rho _*^{-1/2}\, \cdot \, )  -
 \rho _*^{-1/2} (\Delta \rho ^{1/2}_*)]\,  \rho  \label{transform}
\end{equation}
which  directly follows from  Eq. (3), once we set $\rho = \Psi \, \rho _*^{1/2}$,  see also \cite{brockmann}.

This formula is particularly interesting, since  it  effectively transforms the free noise (Wiener)
generator into  its additive perturbation, e.g.     $\hat{H}$, provided that  $\rho _*^{1/2}$
is known.  This observation will prove  useful in the L\'{e}vy flights context.   In fact Eq. (\ref{transform})
has the structural form  (up to the replacement of $\Delta $ by any L\'{e}vy  noise generator)  of the general transport equation
for $\rho (x,t)$  that encodes a response of non-Gaussian  noise to external potentials, see below.

\subsection{More on $L_{FP} \leftrightarrow  \hat{H}$ transformations}

Before passing to non-Gaussian  jump-type processes, few comments should  be made  to  bridge
  notational  and conceptual  incongruences  between  the  well rooted  methodology of the  statistical mechanics approach to random motion,
   \cite{risken} and  math-oriented  relevant references  \cite{vilela2,nelson}.

In the theory of Markovian diffusion processes, physicists pay a  particular attention to the Fokker-Planck equation
  and  thence  $L_{FP}$, while mathematical practitioners advocate an operator $L_{FP}^*= D\Delta + b\nabla $ that
   is Hermitian adjoint to $L_{FP}$.   Actually,  $L_{FP}^*$  is  an infinitesimal  diffusion   generator
     that is directly  involved in the uniqueness proof for the   Markov process in question.
 We note that Eq. (\ref{mapping}) implies  $ - \rho _*^{1/2} L^*_{FP} \rho _*^{-1/2}=  \hat{H}/2mD $

  If we have in hands a transition probability density $p(y,s,x,t)$,   $s\leq t$, a necessary condition
   for the existence of a unique   Markovian dynamics  is that  $p$ solves the   so-called backward  (first Kolmogorov) diffusion  equation
   \begin{equation}
\partial _s p = - D\Delta _y p - b\nabla _y  p= - L^*_{FP}(y,s) p \, ,
\end{equation}
 where  $b= b(y,s)$.  The  very  same transition probability density solves also  the   forward (second Kolmogorov) equation
\begin{equation}
 \partial _t p= D\Delta _x  p   - \nabla (b\, p) =L_{FP}(x,t)\, ,
 \end{equation}
  where  $b=b(x,t)$.
The latter  (forward) equation, once rewritten with respect to a   pdf $\rho (x,t)= \int p(y,s_0,x,t) \rho (y,s_0) dy$, with $s_0\leq t$,
  is  named the   Fokker-Planck equation  by  physics practitioners.

 We indicate that  $p(y,0,x,t) \, \Delta x $ needs to be interpreted as a probability that a particle trajectory
  started from $y$ at time $0$ will reach a $\Delta x$ vicinity of a point $x$ at a time $t>0$,
   e.g. $\int p(y,s,x,t)dx =1$
  for all $s\leq t$ and all $y \in R$.  In the mathematical literature, $p(y,0,x,t)\equiv p(t,y,x)\equiv p_t(x|y)$ would typically appear.

  More formal description of the above  comments,
   that is close to a standard mathematical lore,   looks as follows, \cite{olk}. Given a  Markov transition function
  $p_t(x|y)$ of  a  random  process on $R$.  The generator of the process  $L$    reads
\begin{equation} (Lf)(y)\;=\;\lim_{t\downarrow 0}{\frac{1}{t}}  [\int_{-\infty }^{+\infty }
 p_t(x|y)f(x)dx\;-\;f(y)]\end{equation}
If a  transition
function is stochastically continuous (a priori presumed, see e.g. \cite{olk}),
then the corresponding semigroup $T_t$ is  defined by
\begin{equation} (T_tf)(y)\;=\; \int_{-\infty }^{+\infty }  p_t(x|y)f(x)dx\end{equation}
and is strongly continuous,  so  that  its generator $L$ is
densely defined in a suitable domain. \\
In such a case we can also define an
adjoint semigroup $T_t^*$ acting on the space of
(probability) densities $L^1(R,\,dx)$,
\begin{equation}  (T_t^*\rho)(x)\;=\;\int_{-\infty }^{+\infty } p_t(x|y)\rho(y)dy\end{equation}
We denote  its generator by $L^*$.  In case of  diffusion-type processes we obviously have the generic outcome
 $L^*= L_{FP}$ and $L = L^*_{FP}$. \\
 All that may be checked by inspection in case of the familiar Ornstein-Uhlenbeck  (OU) process, \cite{nelson}.
We scale away physically relevant constants  taking  $L= b(y \nabla _y +  (1/2) \Delta _y $
as a process generator,   where  $b(y)= -y$.
 The OU     transition density  reads
   \begin{equation}
 p(y,0,x,t) \equiv  p_t(x|y)  = [\pi  (1- e^{-2 t})]^{-1/2}\,
 \exp \left[ -  {\frac{(x- e^{- t}y)^2}{(1- e^{-2 t})}}\right]
  \end{equation}
Here $L^* = (1/2)\Delta _x  - \nabla _x [b(x) \, \cdot ]$   and $b(x)= -x$.
This process is  stationary (time-homogeneous), hence we can  safely  replace  $t$  by   $(t-s)\geq 0$, thus passing  to
 $p(y,s,x,t)$  and the related forward and backward   (Kolmogorov)  equations.\\

{\bf Remark }:
 One may  explicitly  resort to  dimensional units, with  $x$   being  replaced by the velocity label $v$, \cite{nelson}.   Then
    the transition probability  density of  the  OU process takes the form
 $$p(u,s,v,t) = (2\pi \beta  D \{1 - \exp [-2\beta  (t-s)]\})^{-1/2}
 \cdot \exp \left(-  {{{ \{v-u \exp[-\beta (t-s)]\}^2}\over
 {2\beta D\{1- \exp[-2\beta (t-s)]\}}}} \right)$$
 with $s<t$  and  has an  asymptotic ($t\rightarrow \infty $) invariant  density  $\rho (v)=
 (2\pi  \beta  D)^{-1/2}\cdot exp(- v^2/ 2\beta  D)$. The drift of the
 process reads $b(v)=-\beta  v$ and $p$ solves the Fokker-Planck
 (second Kolmogorov)  equation
 $\partial _tp= D\triangle _vp - \nabla _v(bp)$.   Here $\beta  $ is a friction coefficient  and $D= k_BT/m\beta $ reflects Einstein's fluctuation-dissipation relationship;   $k_B$ is   the  Boltzmann constant,
   $T$ stands for an equilibrium  temperature of the thermostat.\\

\subsection{Dynamical duality:  "Euclidean" mapping  exemplified}

In the light of  the above  discussion, compare e.g. Eqs. (1) and (3),   it appears   quite  persuasive   to    execute
(at least formally) the Wick rotation
in the complex time plane $it \rightarrow t\geq 0;   \, \, \, \,  \hbar \rightarrow 2mD$,   under  the restriction $t\in R^+$ :
\begin{equation}
  \exp(-i\hat{H}t/\hbar) \psi _0 = \psi _t  \Longrightarrow
\exp(-t\hat{H}/2mD)\Psi _0 = \Psi _t     \label{duality}
\end{equation}
that maps between diffusion-type and quantum mechanical patterns of dynamical behavior, \cite{guerra,guerra1,albeverio,gar0}. In the
absence of external potentials   (free case) we have
$\hat{H} =  -2mD^2 \Delta $.

For clarity of discussion, it is  instructive to invoke explicit  examples.  We pass to  one spatial dimension and    rescale (or completely scale away)
 a diffusion coefficient.
Given a  spectral solution for $\hat{H}= - \Delta + V  \geq 0$   in $L^2(R)$,  the  integral kernel of   $\exp(-t\hat{H})$
reads ($t\rightarrow it$ gives rise to the  kernel of    $\exp(-it\hat{H})$)
 \begin{equation}
k(y,x,t)= k(x,y,t)= \sum_j \exp(- \epsilon _j t) \, \Phi _j(y) \Phi ^*_j(x). \label{spectrum}
\end{equation}

 Remember that we assume  $\epsilon _0=0$  and
the sum may be replaced by an integral in case of a continuous spectrum. Then  one needs to employ  complex-valued  generalized
 eigenfunctions, like e.g. $\Phi _j(x) \rightarrow \Phi _p(x) = (2\pi )^{-1} \,  \exp(ipx)$.
 Indeed, if  we  set   $V(x)=0$ identically,  a familiar heat kernel is  readily obtained :
\begin{equation}
k(y,x,t)= [\exp(t\Delta )](y,x) = {\frac{1}{2\pi }} \int \exp(-p^2t)\, \exp(ip(y-x)\, dp=  \label{heat}
\end{equation}
$$(4\pi t)^{-1/2}\, \exp[-(y-x)^2/4t]\, ,$$
in accordance with $\int \exp(-\sigma ^2 p^2) \exp(-ipx) dx = (\pi /\sigma )^{1/2} \exp( -p^2/4\sigma ^2) $,   where $\sigma >0$.
We note that the kernel of $[\exp(t D \Delta )]$  appears after  changing   the time-scale in (\ref{heat}), $t\rightarrow Dt$.
A formal identification $D\equiv 1/2$  gives   the kernel   that is often met in the mathematical literature and corresponds
to   $(1/2)\Delta $  instead of $\Delta $.

Consider $\hat{H}= (1/2)(-\Delta + x^2  - 1)$ (e.g. the  rescaled  and $(-1)$ renormalized  harmonic oscillator Hamiltonian).
The integral kernel of $\exp(-t\hat{H})$
  is given by a rescaled form of the classic   Mehler formula, \cite{lorinczi,carlitz}:
  \begin{equation}
 k(y,x,t) = [\exp(-t\hat{H})(y,x)=   {\frac{1}{\sqrt{\pi }}} \exp[- (x^2+y^2)/2] \, \sum _{n=0}^{\infty } {\frac{1}{2^n n!}} H_n(y) H_n(x)
\exp(-nt) =
\end{equation}
  $$(\pi [1-\exp(-2t)])^{-1/2} \exp \left[- {\frac{1}2} (x^2-y^2) -    {\frac{(x- e^{- t}y)^2}{(1- e^{-2 t})}}\right] $$
where $\epsilon _n= n$, $\Phi _n(x) = [4^n (n!)^2 \pi ]^{-1/4} \exp(-x^2/2)\, H_n(x)$ is the $L^2(R)$  normalized Hermite  (eigen)function, while
$H_n(x)$ is the n-th Hermite polynomial  $H_n(x) =(-1)^n (\exp x^2) \, {\frac{d^n}{dx^n}} \exp(-x^2) $.

 The normalization condition  $\int k(y,x,t) \exp[(y^2-x^2)/2]\, dy =1$   actually  defines a transition probability density
 of the Ornstein-Uhlenbeck process (see e.g. Eq. (\ref{OU}))
 \begin{equation}
p(y,0,x,t) \equiv p(y,x,t) = k(y,x,t)\, \rho _*^{1/2}(x)/ \rho _*^{1/2}(y)  \label{OU}
\end{equation}
   with $\rho _*(x)=\pi ^{-1/2} \exp(-x^2)$.

 A more familiar form of the  Mehler  kernel reads (note the presence of $\exp(t/2)$ factor)
 \begin{equation}
 k(y,x,t) = {\frac{\exp(t/2)}{(2\pi  \sinh t)^{1/2}}}  \exp  \left[ - {\frac{(x^2+y^2)\cosh t  - 2xy}{2\sinh t }} \right] \, .
\end{equation}
By formally executing  $t\rightarrow it$  one  arrives at the   free Schr\"{o}dinger   propagator
\begin{equation}
K(y,x,t)= [\exp(it\Delta )](y,x) = (2\pi )^{-1/2} \int \exp(-ip^2t)\, \exp(ip(y-x)\, dp=
\end{equation}
$$(4\pi it)^{-1/2}\, \exp[+i(y-x)^2/4t]$$
and likewise, at that of  ($-1$  renormalized)  harmonic oscillator  propagator
\begin{equation}
 K(y,x,t) = {\frac{\exp(it/2)}{(2\pi i \sin t)^{1/2}}}  \exp  \left[ +i {\frac{(x^2+y^2)\cos t  - 2xy}{2\sin t }} \right]
 \end{equation}
 Learn  a standard Euclidean   (field) theory lesson concerning multi-time correlation functions. In the
  exemplary harmonic oscillator case, $t>t'>0$;  $t\rightarrow it$ results in:
\begin{equation}
E[X(t')X(t)] = \int  \rho _*(x')\,   x'\,  p(x',t',x,t)\,  x\,  dx dx'  = (1/2)\, \exp [-(t-t')]   \Longrightarrow
\end{equation}
 $$W(t',t) = \langle \psi _0, \hat{q}_H(t) \hat{q}_H(t')\psi _0\rangle =  (1/2) \exp [-i(t-t')]$$
where $\hat{q}_H(t)$ stands for the position operator in the Heisenberg picture, \cite{vilela2,guerra}.

The major message of our discussion is  that we have  encountered  the dual   dynamical patterns of behavior that  follow equally realistic  clocks.
The Euclidean mapping (Wick rotation) is merely a mathematical artifice  transforming one  dynamical
model into another, \cite{guerra1,gar0,wang}, with an evolution following  a common time scale.  Obviously, once on the quantum  level, we may extend
the validity of the formalism from $t\in R^+$   to all  times $t\in R$.

{\bf Remark 1:} In relation to  the "Euclidean" label, we  may invoke  the statistical physics lore of the $50$-ies and $60$-ties, by  passing   to an integral kernel of the density
operator,  that is parameterized  by equilibrium values of the temperature. To this end one should  set e.g.  $t\equiv \hbar\omega /k_BT$ for a  harmonic  oscillator with a proper   frequency $\omega $ and
 remember about evaluating the normalization factor $1/Z_T$,  where $Z_T$ stands for a partition function of the system.

 {\bf Remark 2:}  In the confined  (due to a suitable choice  of an external potential)  regime,  the    ground state  $\phi $, \,   $\hat{H}\phi = 0$,
 of a   nonlocal   self-adjoint and non-negative  Hamiltonian-type operator $\hat{H}$,     induces   a pdf $|\phi |^2= \rho _*$,
which  actually is an asymptotic  target of  an associated   jump-type stochastic process.
 The  initial  knowledge of the ground state of $\hat{H}$ and the wellposedness  of the affiliated   random motion  problem both   ensure  the existence,
  and  allow  for a  complete   reconstruction of  the   nonlocal   L\'{e}vy-Schr\"{o}dinger  quantum  dynamics and its  dual
   semigroup  partner, see  for example  \cite{vilela1,vilela2,brockmann}.

\section{L\'{e}vy (jump--type)   processes   and  nonlocal random dynamics}

\subsection{L\'{e}vy-Khintchine  formula}

Let us consider a family of  infinitely divisible  probability laws  and affiliated
stochastic Markov processes  $X(t) \rightarrow  \rho (x,t)$ that are characterized  by
 the  celebrated L\'{e}vy-Khintchine  (LK) formula (for any spatial dimension $n\geq 1$)
\begin{equation}
\langle exp[ipX(t)]\rangle := \int_R exp(ipx) \rho (x,t)\,
dx = \exp[-tF(p)]
\end{equation}
where
\begin{equation}
{F(p) = -  \int_{-\infty }^{+\infty } [exp(ipy) - 1 -
{ipy\over {1+y^2}}]
\nu (dy)}  \label{levy}
\end{equation}
and the integral in Eq. (\ref{levy}) is  interpreted  in terms of    its    Cauchy principal value (that is often made explicit by  denoting $\int _{R^n\backslash 0}$
 instead of   the "plain" integral $ \int $);  $\nu (dy)$ stands for
 so-called L\'{e}vy measure  such that
\begin{equation}
\int {\frac{y^2}{(1+y^2)}} \nu (dy)  < \infty
\end{equation}

A   functional form of the L\'{e}vy measure is uniquely singled out by specifying the  L\'{e}vy-Khintchine  exponent $F(p)$ and in reverse.
We shall confine further attention to the family of symmetric stable laws associated with $F_{\mu}(p)=|p|^{\mu }$ where $\mu \in (0,2)$
 and   $F_m(p) =\sqrt {p^2 + m^2} - m, m>0$. Here,  to eliminate inessential dimensional  parameters,  we have
chosen suitable units.
 We note that   we have in fact    encountered   a  re-scaled version of the  classical
 relativistic Hamiltonian, which  is   better known in  the   dimensional  form
  $\sqrt {m^2c^4+c^2p^2 }-mc^2$, where $c$ is the velocity of light.

Its $m\rightarrow 0$ limit is well defined and  equals
$c|p|$  which    directly   refers to the Cauchy (symmetric stable  $\mu =1$) probability law.
 We point out as well that Eq. (\ref{levy}) represents   a reduction  of the complete LK formula  to the  integral  term which
  determines    jump-type probability  laws and processes. We have disregarded a  Gaussian contribution of the  form  $F(p)=p^2/2$,
  identifying  the Wiener noise and process,
  whose properties  underly the reasoning of Section II.

To see the problem from a broader perspective,
we recall that a characteristic function of a random variable $X$  completely determines a probability distribution of that variable.
 If this distribution admits a density we can write $E[\exp(ipX)] =  \int_R  \rho (x)  \exp(ipx) dx$.
Infinitely divisible probability laws  are  classified  by   the general
 L\'{e}vy-Khintchine formula
\begin{equation}
E[\exp(ipX)] =  \exp[-F(p)] =  \exp \{ i\alpha p - (\sigma ^2/2)p^2  +  \int_{-\infty }^{+\infty } [exp(ipy) - 1 -
{\frac{ipy}{1+y^2}}] \nu (dy)\}    \label{LK}
\end{equation}
where $\nu (dy)$ stands for the  L\'{e}vy measure.

To elucidate the meaning of the deterministic term, we  invoke
 the Cauchy probability density function in the form $\rho (x)= (\sigma /\pi ) [(x-\alpha )^2 +\sigma ^2]^{-1}$.
 Its  characteristic function reads   $E[\exp(ipX)]= \exp(i\alpha p - \sigma |p|)$ and the term $- \sigma |p|$ can be verified to
  come out  from the  (scaled by $\sigma $) integral contribution in Eq.~(\ref{LK}).

 A direct consequence of an infinite divisibility of the considered probability laws is that the
"free" noise  fully determines a corresponding  Markov process   through
\begin{equation}
E[\exp(ipX_t)]= \exp[-t F(p)] \, .
\end{equation}
Accordingly, any composite "free" Markov process $X(t)=X_t$  can be interpreted  (decomposed) as follows:
 $X_t = \alpha t  + \sigma B_t + J_t + M_t$   where $B_t$ stands for the free Brownian motion (Wiener process), $J_t$ is a Poisson process
while $M_t$ is a general jump-type process
(more technically, martingale with jumps).

In particular, by disregarding the deterministic and jump-type   contributions in the above, we are left with the Wiener noise $X= \sigma B$.
For a  Gaussian  pdf  $\rho  (x)= (2\pi \sigma ^2)^{-1/2} \exp (-x^2/2\sigma ^2)$ we directly evaluate
$E[\exp(ipX)]= \exp(-\sigma ^2 p^2/2)$.  It is enough to  set $\sigma ^2 = 2Dt$ to   recover the induced Wiener   process with
 $E[\exp(ipX(t)]= \exp(-tDp^2)$ and  $F(p)= Dp^2$.

\subsection{Canonical quantization}

 At this point we  employ  a substitution procedure, that is in fact  a canonical quantization step,
 up to the explicit  presence  of $\hbar$:
 \begin{equation}
  p \rightarrow \hat{p}=-i \nabla  \Longrightarrow   F(p)\rightarrow F(\hat{p})
 \, . \label{quant}
 \end{equation}
We point out that, in view  the standard Fourier representation in use,
a casual quantum  mechanical operator  notion  $(\hat{x}f)(x)= xf(x)$ is implicit.
No covariant position operator, like  that of the Newton-Wigner type, \cite{barut}, is here-by addressed.

The domain of   $F(\hat{p})$  is tacitly  placed in
$L^2(R^n)$, whose element  is in fact   any considered $\rho (x)$.
 That, irrespective of its $L^1(R^n)$  normalization.
  A direct consequence of   the $L^2(R^n)$ arena  choice  for  the action of all considered  operators is
  that  the Fourier transform is a unitary operation.

  This  is sufficient for the validity of  standard
  position-momentum  type uncertainty relations, since  they  automatically follow from  the properties of the direct and
  inverse Fourier transforms. This should be compared with Refs. \cite{hardy,folland},  where Hardy's theorem concerning an interplay between the  localization
  of a function and that of its Fourier transform   has been analyzed  (a folk transcript reads: a function and its Fourier transform cannot
   both be both very small).  Here we also refer to Ref. \cite{wiese} for  a discussion of
   uncertainty relation in relativistic quantum mechanics and  Ref. \cite{romp,stat} for
    a complementary   view upon  uncertainty relations in  random motion. See also \cite{ibb1,ibb2}.

 If $F(p)$  is given in the L\'{e}vy-Khintchine form  (\ref{LK}), then   $\exp [-tF(\hat{p})]$  is   a   contractive semigroup operator.
 In particular this pertains to  $\exp(tD\Delta )$ where   $\Delta  $  denotes the spatial  Laplacian and   the Hamiltonian
  $ (1/2mD) \hat{H}= F(\hat{p})= D\hat{p}^2= -D\Delta $,   has been   previously   employed in the discussions of Section II.
Note that we can get rid of the constant $D$ by rescaling the time parameter.

From now on we shall restrict considerations to symmetric probability distributions  of  pure random jumps. The (infinitesimal)
 L\'{e}vy measures $\nu (dy)$ are odd with respect to the spatial inversion, e.g. $\nu (-dy) = - \nu (dy)$.
Therefore, the regular form of the L\'{e}vy-Khintchine formula (\ref{LK}), while reduced to the integral expression (\ref{levy}),  becomes  further
 simplified (Cauchy principal value of the integral is implicit, see e.g. \cite{gar,olk,cufaro}):
\begin{equation}
{F(p) = -  \int_{-\infty }^{+\infty } [exp(ipy) - 1]
\nu (dy)}  \label{mu}
\end{equation}
which in view of the canonical quantization recipe (\ref{quant}) defines the action of the semigroup generator $- F(\hat{p})$ on
 functions in its domain according to:
\begin{equation}
- F(\hat{p})f(x) =  + \int [f(x+y)-f(x)] d\nu(y)\, . \label{comment}
 \end{equation}
We emphasize that a generically singular behavior of the L\'{e}vy measure in the vicinity of zero needs the  (counter)term containing $-f(x)$ for consistency
reasons. There is no clean  way to eliminate this contribution from relevant  integral formulas.
 This prohibitive statement should be kept in memory while
getting through a number of physics-oriented papers on related topics, as examples see e.g.  formulas (2.11) through (2.26) in \cite{remb}
 or (14), (15) in   Ref. \cite{horwitz}, c.f.  the   next subsection.

As mentioned before, in the family  of infinitely divisible probability laws our attention is focused on  symmetric stable laws associated with
$F_{\mu}(p)=|p|^{\mu }$ where $\mu \in (0,2)$
 and   $F_m(p) =\sqrt {p^2 + m^2} - m$ with $m\geq 0$.   Since  many of our arguments will not rely on the specific spatial dimension
  we shall  reproduce $n$-dimensional versions of the corresponding L\'{e}vy measures.
   In particular,  the $\mu \in (0,2)$-stable  Hamiltonian  operator  $ \hat{H}_{\mu } = F_{\mu}(\hat{p})= |\Delta |^{\mu /2}$,
    induced by $F_{\mu}(p)=|p|^{\mu }$,
acts upon functions in its domain as follows   \cite{bogdan,ryznar}:
\begin{equation}
- |\Delta |^{\mu /2} f(x)  =   \int [f(x+y) - f(x)]\,
\nu _{\mu }(dy) =
{\frac{2^{\mu }\, \Gamma ({\frac{\mu +n}{2}})}{\pi ^{n/2}|\Gamma (- {\frac{\mu }{2}})|}}
\, \int {\frac{f(y)-f(x)}{|x-y|^{\mu +n}}}\, d^ny   \label{free1}
\end{equation}
 where $x\in R^n$  and  $\nu _{\mu }(dy)$ stands for a (self-defining) L\'{e}vy measure
  $\sim 1/|y|^{\mu  + n}$. All  above  integrations are  understood in the sense  of the Cauchy principal value.

  Here, $ -|\Delta |^{\mu /2} $   is a nonlocal   (pseudo-differential, fractional)  generalization of the
 ordinary Laplacian  $\Delta $. It is associated with symmetric stable probability laws and related jump-type  stochastic processes (here
 called L\'{e}vy flights),  \cite{vilela2,brockmann,olk1}-\cite{olk}. We note that the coefficient in Eq.~(\ref{free1}) has been chosen  to secure
 $\int_{R^n} [1- cos(p y)]d\nu _{\mu } (y)= |p|^{\mu }$, \cite{bogdan}, compare e.g. also Eq. (\ref{mu}).

The relativistic (named also quasi-, semi-,  pseudo-relativistic or Salpeter)  Hamiltonian operator
  $\hat{H}_m = F_m(\hat{p})=[\sqrt {- \Delta + m^2} - m] $,  induced by
 $F_m(p) =\sqrt{p^2 + m^2} - m, m\geq 0$,
acts in its domain according to  \cite{ichinose}:
\begin{equation}
- [\sqrt {- \Delta + m^2} - m] f(x)  =  \int [f(x+y) - f(x)]\,
\nu _{m}(dy) =
2 \left( {\frac{m}{2\pi }}\right)^{(n+1)/2}    \int [f(y)-f(x)]
 {\frac{K_{(n+1)/2}(m|x - y|)}{|x- y|^{n+1}}}\,  d^ny \, .  \label{quasi}
\end{equation}
It is  instructive to notice that  the $m\rightarrow 0$  limit of the relativistic semigroup generator exists and coincides with   $\mu =1$ (Cauchy)  stable
 generator  $- |\Delta |$.  In the above $K_{\nu }(z)$    is the modified Bessel function of the third  kind of  order  $\nu  $.\\

\subsection{L\'{e}vy flights: transition pdfs and transport equations.}

"Free" semigroup kernels coincide with transition probability densities of the pertinent jump-type processes. As a complement to  definitions of the previous
two subsections, let us  introduce
 calculational tools that  are based on Fourier transformation techniques.
   An advantage of  a functional analytic lore is that  contractive semigroup operators, their generators and  the
   pertinent integral  kernels can be directly deduced from  the L\'{e}vy-Khintchine formula, see also  \cite{gar}.

We extract  from  semigroup generators  the  "free" Hamiltonians, defined  up to dimensional constants which can be easily recovered if needed. They have
 the form $\hat{H}_0=F(\hat{p})$, where  $\hat{p}=-i \nabla $ stands for the momentum  operator  (here we put  $\hbar =2mD =1$).
Let $f(x)$ be function in the domain of $F(\hat{p})$  and $\tilde{f}(p) =(2\pi )^{-n/2} \int f(x)\exp(-ipx)\, d^nx $  its    Fourier transform.
We define
\begin{equation}
[\exp(-t  \hat{H}_0]f(x)   =   (2\pi )^{-n/2}   \int_{-\infty }^{+\infty } \exp(-tF(k)) \exp(ikx)
\tilde{f}(k) d^nk = [\exp(-tF(p)) \tilde{f}(p)]^{\vee }(x) \label{spectrum1}
\end{equation}
where the superscript $\vee $ denotes the inverse Fourier transform.
Let us set
\begin{equation}
k_t=(2\pi )^{-n/2}  [\exp(-tF(p)]^{\vee }\, .  \label{kernel}
\end{equation}
Then the
action of $\exp(-t \hat{H}_0)$ can be given in terms of a convolution (i.e. by means of an integral kernel $k_t\equiv  k(x-y,t)=k(y,0,x,t) $):
\begin{equation}
\exp(-t\hat{H}_0)f = [\\exp(-tF(p)) \tilde{f} (p)]^{\vee } = f*k_t    \label{conv}
\end{equation}
where $ (f*g)(x): =\int  g(x-z)f(z)dz  $  is a convolution of  functions $f$ and $g$.

In view of  Eq.~(\ref{kernel}) we have in hands   Fourier integral redefinitions  for all pertinent semigroup kernels.
   We note  that  those kernels  play the  role  of  transition probability densities of the underlying  Markovian  jump-type processes.
In particular, $k_{\mu }(x-y,t-s)$ has the form
\begin{equation}
k_{\mu }(x,t) = (2\pi )^{-n}  \int \exp(-ipx - t|p|^{\mu })\, d^np \label{stable}
\end{equation}
while $k_m(x-y,t-s)$ reads
\begin{equation}
 k_m(x,t) = (2\pi )^{-n}  \int \exp[-ipx - t  (\sqrt{m^2 + p^2} - m)]\, d^np\, . \label{relativistic}
 \end{equation}
For completeness  one should recall a  Fourier form of the heat kernel  $k(x-y,t-s)$, compare e.g.  Eq.~(\ref{heat}).
\begin{equation}
k(x,t)= [\exp(t\Delta )](x) =(2\pi )^{-n} \int  \exp(-ipx   - tp^2)\, d^np \, ,
\end{equation}
All the  above (generalized) heat kernel formulas stem from a  spectral resolution of the corresponding (rescaled) Hamiltonian operator, compare e.g.
our  discussion of  Eqs. (\ref{spectrum}),    (\ref{heat})  and (\ref{spectrum1}).

 One should be aware that closed analytic outcomes of  Fourier integrals are scarce in the present framework. We shall reproduce explicit
 formulas for integral kernels of the Cauchy and quasi-relativistic semigroups, see e.g. \cite{lorinczi,ichinose,ichinose1}.
Namely, upon reintroducing  suitable    physical constants  ($\hbar $ is kept equal 1, see however \cite{remb}),  the  integral kernel of the  semigroup operator
$\exp[-t(\sqrt{-c^2\Delta +m^2c^4} - mc^2)]$  reads:
\begin{equation}
 k_m(x,t)= 2 \left({\frac{m}{2\pi }}\right)^{(n+1)/2} c^{(n+3)/2} [t\,
 \exp(mc^2t)]  (x^2+c^2t^2)^{-(n+1)/4}\, K_{(n+1)/2}(mc\sqrt{x^2 +c^2t^2}) \, . \label{k}
\end{equation}
In view of  $t^{-1} k_m(y,t) dy \rightarrow \nu_m(dy)$as $t\downarrow 0$,  after putting $c\equiv 1$, we retrieve the previous formula
 for the L\'{e}vy measure as employed in Eq. (\ref{quasi}).

 The mass $m\rightarrow 0$ limit of  $k_m(x,t)$ does exist and equals  to the Cauchy kernel $k_{0}(x,t)$:
\begin{equation}
k_0(x,t) =   \Gamma \left({\frac{n+1}{2}}\right)  {\frac{ct}{[\pi (x^2+c^2t^2)]^{(n+1)/2}}}  \label{cauchy}
\end{equation}
whose more popular form  (in view of computational simplicity) employs  $c\equiv 1$.

In view of the semigroup dynamics   rule   $[\exp(- t\hat{H})f](x)= f(x,t)$,
we readily get transport equations  for pdfs driven by "free" noise.  The equations
\begin{equation}
\partial _t \rho (x,t) = - |\Delta |^{\mu /2}\rho(x,t)=   - |\nabla |^{\mu  }  \rho = \int [\rho (x+y,t) - \rho (x,t)]\,
\nu _{\mu }(dy)
 \end{equation}
where $\mu \in (0,2)$  and
 \begin{equation}
  \partial _t \rho (x,t)  = - [\sqrt {- \Delta + m^2} - m]  \rho (x,t)  = \int [\rho (x+y,t) - \rho (x,t)]\,
\nu _{m}(dy)  \label{quasi1}
  \end{equation}
   are so-called master equations for  the pertinent    jump-type processes and, in the present setting,    replace the standard  Fokker-Planck type equation,
   appropriate for a diffusive transport.
     We recall  that  the Cauchy principal value need to  be attributed to  the  integrals involved and  one  \it must not  \rm disregard the $\sim \rho (x) $ counter term in those formulas.

 As an exemplary formula we   reproduce  the master equation for the 1D  Cauchy transition probability density $k_{1}(x-y,t-s)  =  p(y,s,x,t)$ with $s<t$:
   \begin{equation}
   \partial _tp(y,s,x,t)  = \int [p(y,s,x+z,t)  -  p(y,s,x,t)] d\nu _1(z) =  \int  {\frac{p(y,s,z,t)   -  p(y,s,x,t)}{\pi (x-z)^2}} \, dz
   \end{equation}

\subsection{Continuity equation  $\partial _t \rho = - \nabla j$  for  symmetric    stable processes.}

\subsubsection{Inversion   of   $\nabla $.}

In case of the Brownian motion (Wiener process) the continuity equation $\partial _t \rho = - \nabla j$ is
merely another form of the heat equation $\partial _t \rho = D\Delta \rho $, where a definition  of
$j = v \rho $ along  with   $v = -D\nabla \ln \rho $ does the job.
Things become less obvious if the free transport equation is to take  the L\'{e}vy  form  appropriate for a symmetric stable process.

For concreteness, instead of    $\partial _t \rho =
 - \gamma |\nabla |^{\mu } \rho = -  \gamma  |\Delta |^{\mu /2}  \rho = - \gamma  \nabla j$,  with $\mu \in (0,2)$ and
 $\gamma $ being the jump intensity parameter, let  us  discuss  the Cauchy   case $\mu =1$  in  some  detail.
We note that $\sqrt {m^2c^4+c^2p^2 }-mc^2$ in the mass $m=0$ limit reduces to $c|p|$. Accordingly we may  pass to
the nonlocal equation  $\partial _t\rho  = - c |\nabla | \rho $, c.f. Eq. (\ref{stable})  which has
a classic    Cauchy  solution
\begin{equation}
\rho (x,t) =  k_1(x,t)= {\frac{1}{2\pi  }}  \int _{-\infty } ^{+\infty }  \exp(-ikx  - |k| \lambda  ) \, dk =
 {\frac{1}{\pi }} {\frac{\lambda }{\lambda ^2 + x^2}}
\end{equation}
with $\lambda = \lambda (t) = b+ct$, $b\geq 0$.
We note also that
\begin{equation}
|\nabla | \rho = {\frac{1}{2\pi  }}  \int _{-\infty } ^{+\infty } |k|   \exp(-ikx  - |k| \lambda  ) \, dk =
 {\frac{\lambda ^2-x^2}{\pi (\lambda ^2 + x^2)^2}}~\label{60}
\end{equation}
Accordingly   $\partial _t \rho =  - c |\nabla | \rho = - \nabla j  $,  where an  analytic  expression for the probability current $j(x,t)$
 is given  in terms of  an indefinite integral (a symbolic  inversion  $\nabla ^{-1}$ of $\nabla $):
\begin{equation}
j(x,t) =   - \int [\partial _t \rho (x,t)]\, dx  =    {\frac{c}{\pi }} {\frac{x}{x^2 + \lambda ^2}}  =   v\, \rho \, . \label{inversion}
\end{equation}
Here   $v(x,t)= j/ \rho =  c\, x/ \lambda $  and we  formally have $c  \nabla ^{-1} |\nabla | \rho = j$.
In view of $\lambda = b+ct$,  for sufficiently  large times  $v \sim   x/t$ holds true.
We  can  also  write $v=  -(c/2\lambda ) \nabla \ln \rho $.
This should  be compared with the  Brownian outcome
 $v =-D \nabla \ln \rho  \sim +x/t$   (to this end set e.g.   $\rho \sim \exp(- x^2/2Dt)$).

The inversion formula $\nabla ^{-1}  \nabla j  \rightarrow j$   reported in Eq.~(\ref{inversion})  may  be interpreted  in terms of the Fourier representation.
Namely,   the Fourier transform $\tilde{j}(p)$ of   $j(x,t)$  reads
\begin{equation}
\tilde{j}(k,t) = {\frac{1}{\sqrt{2\pi }}}  \int \exp (-ikx)   j(x,t)dx  =  -i  sgn(k)\,  {\frac{c}{\sqrt{2\pi }}} \exp(-\lambda |k|) \, .
\end{equation}
We define the action of $\nabla ^{-1}$ through (cannot be extended beyond 1D):
\begin{equation}
\nabla ^{-1}  \exp(-ikx) =    {\frac{1}{(- i k)}}  \exp(-ikx) \, . \label{inv}
\end{equation}
Accordingly, we have from Eq.~(\ref{60})  ($|k|/k = sgn (k)$, next employ $k=-p$)
\begin{equation}
c  \nabla ^{-1} |\nabla | \rho (x,t) =  {\frac{i c}{2\pi  }}  \int _{-\infty } ^{+\infty } sgn(k)   \exp(-ikx  - |k| \lambda  ) \, dk  =
 {\frac{1}{\sqrt{2\pi } }}  \int _{-\infty } ^{+\infty } \tilde{j}(p,t) \exp(ipx) dp = j(x,t)
\end{equation}
One must be aware that  rather  restrictive assumptions are necessary for a function $g$ to belong to the domain of $\nabla  ^{-1}$.   In fact we can use   $\nabla  ^{-1}$ only when  $g$ is "sufficiently nice".
Such  phrase  is   sometimes used in the mathematical  literature in the analogous context   of properly defining $|\Delta |^{-\mu /2}$ as  a legitimate operator, \cite{hu}. As shown above, a function $g= \nabla f $  where
$f$ is sufficiently fast decreasing  at infinities,  is an example of a "sufficiently nice"  function.

\subsubsection{Inversion    of   $|\Delta |^{\mu /2}$.}

One of our motivations is to analyze the origin and meaning   of   formulas of the type $(-\Delta  )^{\mu /2  -1} \nabla \Phi $
appearing in the literature  devoted to so-called fractional quantum mechanics,   also in relation to the probability current concept, \cite{laskin,laskin1,dong}. There-in,   under the restriction $1<\mu <2$, hence
  excluding the previously considered case of $\mu =1$ and generally $\mu \in (0,1)$.  Another motivation
   is related with the Landau-Peierls  notion of the photon wave function, the definition of
   which employs  $(-\Delta  )^{-1/4}$,  see e.g. \cite{landau,good,ibb,hawton}. We note in passing that  those operators per se are not strictly positive, but  merely
    non-negative. Therefore suitable domain restrictions need to be carefully observed (e.g. this point is not a  problem
     in the photon wave function  construction, where  $p=0$ is excluded by the transversality condition, see e.g. \cite{ibb,ibb1}.

Arguments of the present section  are  not necessarily  limited by the dimensionality of space ($D\geq 1$).
Our discussion is based on the Fourier representation, hence we assume a priori the existence of  (both direct and inverse)
 Fourier transforms   of considered functions
(the  inverse transform is  not  a mathematically obvious notion,  albeit
 always  tacitly assumed  to hold true  in the physics  literature).
  We indicate how the domain of  an  inverse fractional Laplacian    may be interpreted, see e.g. also \cite{hu}.

The action of any $|\Delta |^{\mu /2}$  upon the Fourier transformable function $f(x)$  can be defined as a multiplicative modification
 $|k|^{\mu /2} \tilde{f}(k)$    of   its Fourier transform  $ \tilde{f}(k)$.   Since the inverse Fourier transform
$[|k|^{\mu /2} \tilde{f}(k)]^{\vee }(x) =  g(x)$  is assumed to exist, we have in hands  an example of  a "nice" function.
For such function we can readily   define  the action of   $|\Delta |^{- \mu /2}$  as    $|k|^{- \mu /2} \tilde{g}(k)$  where $\tilde{g}$
stands for the Fourier transform of a  "nice"   function $g$.  In general,  $g$ cannot be arbitrarily chosen  and  rather severe  restrictions
need to  be  observed, to  secure the existence of the inverse   Fourier transform  of  $ |k|^{-\mu /2} \tilde{g}(k)$.
Our ultimate conclusion is that an identity $|\Delta |^{-\mu /2} |\Delta |^{\mu /2}\, f = f$ surely  makes sense for suitable functions $f=f(x)$.

We are now ready to address the meaning of  $\nabla ^{-1}|\Delta |^{\mu /2}$ necessarily appearing in the transformation of
$\partial _t\rho = - |\Delta |^{\mu /2} \rho $  into a continuity equation   $\partial _t \rho = - \nabla j$.
We consider $k_{\mu }(x,t) = \rho _{\mu }(x,t)$, see Eq. (\ref{stable}). For clarity of discussion we pass to  the space dimension   $D=1$.
Accordingly
\begin{equation}
\nabla ^{-1}|\Delta |^{\mu /2}  \rho _{\mu }(x,t) = {\frac{i }{2\pi  }}  \int _{-\infty } ^{+\infty } sgn(k) \, |k|^{\mu  -1}\,
  \exp(-ikx  - |k|^{\mu } \lambda  ) \, dk
\end{equation}
where the sign  factor $sgn(k)$ is a crucial ingredient that makes a difference between $\nabla ^{-1}|\Delta |^{\mu /2} $ and $|\Delta |^{(\mu -1)/2}$.

We  can be sure of the existence of the above Fourier resolution for $j(x,t)$  only in the  stability parameter range $\mu \in [1,2)$.
For $\mu \in (0,1)$,  the function $\partial _t\rho _{\mu}(x,t)$ cannot be represented as a gradient function.
 Therefore,  for $\mu \in (0,1)$,  the
 transport equation $\partial _t\rho _{\mu }(x,t) = - \Delta ^{\mu /2} \rho _{\mu }(x,t)$
 cannot be  given  a  functional form of the  continuity equation.

\subsection{Pdf transport   equations induced by confining  L\'{e}vy semigroups}

 In case of   "free" semigroups, there is no asymptotic invariant densities so that similiar to  the  Brownian  case, c.f.
    $\partial _t  \rho = + \Delta \rho $      (we disregard so far   irrelevant coefficients),
we deal with a sweeping motion.

The situation changes drastically, if we pass to  the  confining regime   of a jump-type process, with an invariant density  $\rho _*$.
It is rather obvious, \cite{brockmann,gar1,gar2}, that by choosing   $\rho _*^{1/2}$ and  then making
  a formal replacement of $(- \Delta )$ by $|\Delta |^{\mu /2}$  everywhere
in  Eq. (\ref{transform}), followed by an adjustment of  noise  intensity parameters $D\rightarrow \lambda $ we end up with:
\begin{equation}
\partial _t \rho =  \lambda  [-  \rho ^{1/2}_*|\Delta |^{\mu /2}(\rho _*^{-1/2}\, \cdot \, )  +
 \rho _*^{-1/2} (|\Delta |^{\mu /2}\rho ^{1/2}_*)]\,  \rho   \, . \label{fractional}
\end{equation}
We  note that, instead of  a specific choice  $|\Delta |^{\mu /2}$, we can equally well employ another   admissible $F(\hat{p})$, like e.g.
$\sqrt {- \Delta + m^2} - m$.  Our arguments extend to those cases as well.

This time evolution of the pdf $\rho (x,t)$  is    formally  induced by the  (would-be)
 L\'{e}vy-Schr\"{o}dinger semigroup $\exp(-t \hat{H}_{\mu })$.

At this point   we  further extend  an applicability of our arguments by considering  a fractional  (pseudo-differential)
generalization   of the  "free"  Hamiltonian  (1)  (set  e.g.  $- D \Delta  \rightarrow      \lambda |\Delta |^{\mu /2}$
 and $V/2mD \rightarrow {\cal{V}}$)   to that with a confining potential:
\begin{equation}
\hat{H}_{\mu }  \equiv  \lambda |\Delta |^{\mu /2} +  {\cal{V}}(x)\, .
\end{equation}
A  tacitly  presumed  ground state $\rho ^{1/2}_*$  corresponds  to its bottom eigenvalue $0$, i.e.  $\hat{H}_{\mu } \rho _*^{1/2}=0$ holds true.
Accordingly,
\begin{equation}
 \partial _t \Psi = - \hat{H}_{\mu } \Psi  =   -  \lambda |\Delta |^{\mu /2} \Psi  -  {\cal{V}}(x)\Psi \, , \label{jump}
\end{equation}
  where
$\Psi (x,t) = \rho (x,t)\, \rho _*^{-1/2}(x)$. A  compatibility condition,
\begin{equation}
{\cal{V}}  =   -\lambda\,  {\frac{|\Delta |^{\mu /2}  \rho ^{1/2}_*}{\rho ^{1/2}_*}} \, , \label{compatible}
\end{equation}
determines ${\cal{V}}$ and  that  in turn allows to expect  an equivalence between  Eq. (\ref{fractional}) and
 Eq. (\ref{jump}).

 There are however some  jeopardies in this formal procedure. An independent
 check is necessary of whether $\hat{H}\geq 0$ actually is  a well defined self-adjoint operator with
  a dense domain of definition, which may not be the case.

Things are much easier, if we first choose an appropriate  $\hat{H}$,  together with its bottom
 eigenvalue $0$  corresponding to the
(ground)  eigenstate $\rho  ^{1/2}_*$. Then the semigroup is given a priori.  Consequently,
 Eq. (\ref{fractional})follows  along  with a proper large-time  asymptotic behavior
  of $\rho(x,t)\rightarrow \rho _*(x)$.

One can rewrite the transport equation $(\ref{fractional}$) in the  canonical
 form  of so-called master equation,  appropriate
for jump-type processes,\cite{gar1,gar2} :
\begin{equation}
\partial _t \rho = \int [ w(x|z)\rho (z) - w(z|x)\rho (x)] dz \, .
\end{equation}
If we replace  the symmetric   jump rate,   like e.g.  (1D stable case is exemplified for clarity of arguments)
\begin{equation} \label{unn1}
w(x|y)= w(x|y)  \sim 1/|x-y|^{1+\mu }
\end{equation}
  by  a non-symmetric  expression
\begin{equation} \label{unn2}
w_{\phi }(x|y)\sim {\frac{\exp [\Phi (x) - \Phi (y)]}{|x-y|^{1+\mu }}}
\end{equation}
 then  $ |\Delta |^{\mu /2} \to  |\Delta |^{\mu /2}_{\Phi }$ and
the corresponding transport equation reads:
\begin{eqnarray}
&&(1/\lambda )\partial _t \rho = - |\Delta |^{\mu /2}_{\Phi }f =
-  \exp (\Phi )\, |\Delta |^{\mu /2}[ \exp(-\Phi ) \rho ]   \nonumber \\
&& +\rho \exp (-\Phi ) |\Delta |^{\mu /2} \exp(\Phi ) \, .\label{unn3}
\end{eqnarray}
We can always  always select
$\exp (\Phi (x)) = \rho _*^{1/2} $ in the above, so arriving at Eq. (\ref{fractional}).\\

{\bf Remark:} Independently of the  method   adopted,  an important difference needs to be spelled out,
 if compared with
the diffusion-type reasoning. Namely, the transport equation  Eq. (\ref{fractional})
{\it cannot} be converted to the   Fokker-Planck form
\begin{equation}
\partial _t\rho = -\nabla \left(- {\frac{\nabla U }{m\beta }}\, \rho  \right) - \lambda |\Delta |^{\mu /2}\rho  \, ,
 \end{equation}
traditionally attributed in the physics literature to L\'{e}vy flights in external conservative force fields ($U(x) $ stands for the
 Newtonian potential, \cite{fogedby,chechkin,dubkov}.  (Note a replacement $- D\Delta  \rightarrow  \lambda |\Delta |^{\mu /2}$
 in $\partial _t\rho =  - \nabla (b\, \rho )+ D\Delta \, \rho $.)
 This  specific  dynamical   inequivalence issue    has been pointed out in \cite{brockmann,gar1,gar2}, see e.g. also \cite{olk,olk1}.

\section{Nonlocal quantum dynamics.}

\subsection{Duality transformation}

Fractional Hamiltonians  $\hat{H}_{\mu }=  \gamma |\Delta |^{\mu /2}$  with  $0 <\mu <2$ and $\gamma >0$ and likewise the quasi-relativistic
one $\hat{H}_m = \sqrt{p^2 + m^2} - m$ with $m\geq 0$ are self-adjoint operators  in
suitable  $L^2(R)$ domains. They are also  nonnegative  operators, so that  the respective fractional  semigroups    are holomorphic
(also named analytic),   and    we can replace the time parameter $t>0$ by a complex one $\sigma =t+is,\, Re \sigma >0$. Accordingly,
a holomorphic extension of  the  L\'{e}vy-Schr\"{o}dinger  semigroup is defined as follows,
 c.f. Eq.~(\ref{spectrum1}):
\begin{equation}
{[\exp(-\sigma \hat{H})]f = [(\tilde{f}\exp(-\sigma F)]^{\vee } = f*k_{\sigma }}\, .
\end{equation}

Here, the integral  kernel reads $k_{\sigma }={1
\over {\sqrt {2\pi }}}[\exp(-\sigma F)]^{\vee }$. Since $\hat{H}$ is
selfadjoint, the
limit $t\downarrow 0$ leaves us with the unitary group $\exp(-is\hat{H})$,
acting in
the same way: $[exp(-is\hat{H})]f = [\tilde{f} exp(-isF)]^{\vee }$, except
that now
$k_{is}: = {1\over {\sqrt {2\pi}}}[exp(-isF)]^{\vee }$  no longer  is
a   probability  measure (transition probability density).

 In view of the unitarity, the unit ball in $L^2$ is an
invariant of the dynamics. Hence  probability  densities, in  the  standard  quantum mechanical form $\rho = \psi ^*\, \psi $
 can be associated with
solutions of the  free   fractional (or Salpeter) Schr\"{o}dinger-type  equations:
\begin{equation}
 i\partial _s \psi (x,s) =   \gamma |\Delta |^{\mu /2} \psi (x,s)
\end{equation}
\begin{equation}
 i\partial _s \psi (x,s) =  [\sqrt{   - \Delta + m^2} - m] \psi(x,s)  \label{salp}
\end{equation}
with  initial data  $\psi (x,0)$. Attempts towards formulating   the  so-called  fractional quantum mechanics can be found in Refs.
 \cite{gar,laskin,laskin1,cufaro}.

 We note the nonlocal action of  motion  generators is  somewhat  blurred in the Fourier representation.
The  pertinent spatial nonlocality becomes   obvious  if  the canonical quantization  is  carried out on the level of the
 L\'{e}vy-Khintchine formula, c.f. subsections III.A and III.B.

Guided by  \cite{gar0}  we   identify  the semigroup time label  $t\geq 0$ with the L\'{e}vy-Schr\"{o}dinger
time label, e.g. set $s\rightarrow t$.   All that amounts to the   duality  (Euclidean) mapping  $t\rightarrow it$ of Section
 II.D., c. f. (1) and  (\ref{duality}), which we exemplify for the symmetric stable noise generators:
\begin{equation}
\partial _t\rho  = -  \gamma |\Delta |^{\mu /2} \rho       \longrightarrow   i\partial _t \psi  =   \gamma  |\Delta |^{\mu /2} \psi
\end{equation}

Stable stochastic processes and their quantum counterparts are plagued by a common disease: it is extremely hard,
if possible at all, to produce insightful analytic solutions.  To get a flavor of intricacies and  technical
subtleties involved,  whose neglect  leads to  erroneous formulas (and a danger  of  untrustworthy  physical conclusions),
we have been quite detailed in the  analysis of  L\'{e}vy   dynamical  semigroups and  their unitary (quantum) partners.
Subsequently,   while developing a general theory, we shall pay  special  attention   to the quasirelativistic (Salpeter) equation and its
$m\rightarrow 0$  Cauchy-Schr\"{o}dinger  limit.  For   clarity of arguments (and computational convenience) unwanted parameters (like $\gamma $) will be
scaled away.

\subsection{Fourier representation advantages and  drawbacks}

\subsubsection{L\'{e}vy-stable case}

Equations (\ref{free1}) and (\ref{quasi})  define the action of L\'{e}vy stable and quasirelativistic generators upon functions in their domain.
The involved  integrals   are interpreted in terms of their Cauchy principal values.
  One must as well  keep in mind a   crucial role of  the   $\sim  -f(x)$ counter term in Eqs. (31) through (34).
  Its presence there is indispensable and appears to have  beeen overlooked in quantum mechanically oriented papers, c.f. \cite{horwitz,remb}.

 To explain that issue, we shall first  discuss in some  detail  a
 fractional $0<\mu < 2$  Laplacian  in space dimension $n=1$. Its spatially nonlocal action upon functions  in a suitable Hilbertian  domain
(domain  issues we relegate to the last section of the paper)  reads:
\begin{equation}
(-\Delta )^{\mu /2} f(x) = |\Delta |^{\mu /2}f(x) \equiv |\nabla |^{\mu } f(x)  =
 -{\frac{\Gamma (1+\mu )\, \sin(\pi \mu /2)}{\pi }}   \int {\frac{f(y)-f(x)}{|x-y|^{1+\mu }}}\, dy  \, . \label{free}
\end{equation}

Let us investigate the properties of   $-|\Delta|^{\mu/2} f(x)$ by turning over to the   Fourier image  $\tilde{f}(k)$  of  $f(x)$.
Eq.~(\ref{free})  yields (a formal interchange of integrations is here-by executed):
\begin{eqnarray}
-|\Delta|^{\mu/2}f(x)=\frac{\Gamma(1+\mu)\sin\frac{\pi \mu}{2}}{\pi \sqrt{2\pi}}
\int_{-\infty}^{\infty}\tilde{f}(k)e^{\imath kx}dk \int_{-\infty}^{\infty}\frac{(e^{\imath ky}-1)dy}{|y|^{1+\mu}} .\label{intf1}
\end{eqnarray}
The integral over $dy$, presuming its very  existence (which is not the case for $\mu =1$), can be calculated as follows
\begin{equation}\label{intf2}
 \int_{-\infty}^{\infty}\frac{(e^{\imath ky}-1)dy}{|y|^{1+\mu}}\equiv 2 \int_{0}^{\infty}\frac{(\cos ky -1)dy}{|y|^{1+\mu}}=
 2|k|^\mu\Gamma(-\mu)\cos\frac{\pi \mu}{2}.
\end{equation}
We note the importance of the restriction $\mu \in (0,2)$ and obvious divergence problems to be taken care of: the $\Gamma $ function
is known to have simple poles at  points $0$, $-1$ and $-2$. Therefore, at $\mu =0, 1, 2$,
 the integral \eqref{intf2} is divergent.
 However, irrespective of how close to  $0$, $1$   or  $2$  the label $\mu >0$ is,   the  integral \eqref{intf2} is well defined.\\
 It is  interesting to observe that  the   divergence of the  the Fourier integral, as $\mu $  approaches $0$, $1$   or $2$,  becomes compensated,
  if we substitute  it back to Eq. \eqref{intf1} and next consider  the limiting behavior of the result:
\begin{eqnarray}\label{intf3}
&&-|\Delta|^{\mu/2}f(x)=\frac{2\Gamma(1+\mu)\Gamma(-\mu)\sin\frac{\pi \mu}{2}\cos\frac{\pi \mu}{2}}{\pi \sqrt{2\pi}}
 \int_{-\infty}^{\infty} |k|^\mu \tilde{f}(k) e^{\imath kx}dk=\nonumber \\
 &&=- \frac{1}{\sqrt{2\pi}}\int_{-\infty}^{\infty} |k|^\mu  \tilde{f}(k)   \label{form}
  e^{\imath kx}dk.
 \end{eqnarray}
  Here we employ the identity (Euler's reflection formula), with an obvious reservation that it  becomes {\it invalid } at
   sharp values $\mu = 0, 1, 2$ of $\mu \in [0,2]$,
   while being operational  for all $\mu \in (0,1)\bigcup (1,2)$:
 \begin{equation}\label{mag}
 \Gamma(1+\mu)\Gamma(-\mu)=-\frac{\pi}{\sin \pi \mu}.
 \end{equation}

This observation clearly identifies some  of frequent  misuses of the formalism if  sufficient attention is not paid to potential  obstacles
(like e.g. an interchange of improper integrals or  a neglect of  the  $\Gamma $ function simple  poles along the
negative semi-axis).  For example, the range of validity of the right-hand-side of  Eq.~(\ref{intf3})  goes beyond $\mu \in (0,2)$ and  admits a
 safe  extension to the   boundary values $0$ and $2$, by-passing as well   the  previously raised  problem concerning $\mu =1$.

  Quite apart from   this  appealing outcome, the   {\it primary}  integral representation (\ref{free}) is not valid  \it at \rm  the boundaries of
 the stability interval  and one needs to resort to the remaining    local   terms of the  general  L\'{e}vy-Khintchine formula.  (\ref{LK}).
More than that, if one presumes  the Fourier representation   Eq. (\ref{form})
as a valid definition of how  $|\Delta|^{\mu/2}$  acts upon functions in  its domain, there is now  clean  way to  go  backwards,
 such   that the   {\it primordial} definition   $(\ref{free})$  actually  could have    been  reproduced.

This  obstacle has been  often overlooked in the literature. That  can  be  explicitly seen  in  publications on
 the Salpeter equation and its solutions, \cite{horwitz,remb}. For example, in the  would be   (actually divergent)   integral kernel formulas
  for operators  $|\Delta|^{\mu/2}$  and
 $[\sqrt {-\Delta + m^2} - m]$ respectively, presented  there-in,  the    $\sim - f(x)$ counter term   is conspicuously missing.
 Below we shall  be  more explicit on this  "missing counter term" issue  in the  discussion of the  1D quasirelativistic case with  $m\geq 0$.

\subsubsection{Quasirelativistic   Schr\"{o}dinger (Salpeter) equation and its  $m\rightarrow 0$  limit.}

In the  physical units, the 1D relativistic  (here named quasirelativistic)  free Schr\"odinger equation is commonly considered in the form
\begin{equation}\label{rfse1}
i\hbar \frac{\partial \phi(x,t)}{\partial t}=\sqrt{m^2c^4-\hbar^2c^2\frac{\partial^2}{\partial x^2}}\  \phi(x,t).
\end{equation}
Denoting $\tilde{f}(k)= (2\pi )^{-1/2} \int_{-\infty}^\infty f(x)e^{-ikx}dx$, $f(x)=  (2\pi )^{-1/2} \int_{-\infty}^\infty \tilde{f}(k)e^{ikx}dk$ and
interpreting  the action of the square root operator $\sqrt{m^2c^4-\hbar^2c^2\frac{\partial^2}{\partial x^2}}\  \phi(x,t)$ in
terms of the series expansion, we  readily arrive at  the  following formal  Fourier representation:
\begin{eqnarray} \label{rfse3}
 \sqrt{m^2c^4-\hbar^2c^2\frac{\partial^2}{\partial x^2}}\  \phi(x,t)   =
\frac{mc^2}{\sqrt{2\pi}}\int_{-\infty}^\infty \tilde{\phi }(k,t) dk\ \sqrt{1-\frac{\hbar^2}{m^2c^2}\frac{\partial^2}{\partial x^2}}\
e^{ikx}=\nonumber \\
\frac{mc^2}{\sqrt{2\pi}}\int_{-\infty}^\infty \tilde{\phi }(k,t) dk\ \left[1- \frac{\hbar^2}{m^2c^2} \frac 12 \frac{\partial^2}{\partial x^2}-
\left(\frac{\hbar^2}{m^2c^2}\right)^2\frac 18 \frac{\partial^2}{\partial x^2}-...\right]\ e^{ikx}= \label{rfse4} \\
 \frac{mc^2}{\sqrt{2\pi}}\int_{-\infty}^\infty \tilde{\phi }(k,t) dk\ \sqrt{1+\frac{p^2}{m^2c^2}}\ e^{ikx}  =
 \frac{1}{\hbar\sqrt{2\pi}}\int_{-\infty}^\infty dp\ \sqrt{m^2c^4+p^2c^2}\, \,  e^{ipx/\hbar}
\tilde{\phi }(p,t). \nonumber
\end{eqnarray}
We note that $\hbar /mc$ stands for so-called reduced  Compton wave-length  and  the momentum label has physical dimensions $p=\hbar k$.

Although we have anticipated the existence of the mass $m=0$ limit in the relativistic Hamiltonian context, the above
 derivation of   \eqref{rfse3}  rings warning bells. Indeed, tacitly presuming the  nonrelativistic  regime  $p^2 \ll m^2c^2$  we have  plainly
  expanded $mc^2 \sqrt{1+ (p^2/m^2c^2)}$ in
 \eqref{rfse4} into Taylor series with respect to $\sim p^2/m^2c^2$ and  evidently we are left with    no   room  for $m \rightarrow 0$ therein.

 Nonetheless,    we can safely  put  $m=0$, after the series resummation - in the last entry of the formula
   \eqref{rfse4}-  so arriving at the
  correct  form of the  Fourier image of $|\nabla |$.  Indeed, \cite{lammerzahl}, to this end we should  consider the ultrarelativistic
  regime  with $p^2 \gg m^2$ and make an expansion  of the   $|p|c \, \sqrt{1 + (m^2c^2/p^2)}$   with respect to  $\sim m^2c^2/p^2$.
   Letting $m\rightarrow 0$  becomes a legitimate operation that replaces  $\sqrt{m^2c^4+p^2c^2}$   by   $c|p|$ in the Fourier representation.
    We  refer to the previous subsection for a discussion  of  how $c |\nabla |$ can  in turn  be recovered, c.f. also Eq. (78) below.

  We point out  that the nonrelatvistic limit does  make sense exclusively in the Fourier  representation.
  More than that,     the limit  $m \to 0$ and Fourier imaging of Eq.~\eqref{rfse1} are not interchangeable operations.
  This is an important  subtlety of the mass  $m\geq  0$   quasirelativistic dynamics, quite akin to those
  raised in  relation to
  L\'{e}vy stable Hamiltonians of the previous subsection.

\subsection{On integral  (kernel)     representations  of    quasirelativistic and Cauchy  generators.}

In the present subsection we take under scrutiny a procedure  \cite{horwitz,remb}  of assigning an integral kernel ({\it not}
 a semigroup kernel discussed in Section III)  to the operator $\sqrt{m^2c^4-\hbar^2c^2\frac{\partial^2}{\partial x^2}}$ and likewise to
 $\hbar c |\nabla |$.  An original argument of Ref. \cite{remb} goes as follows: (i) assume (\ref{rfse3}) to hold true,
  (ii) take the inverse Fourier transform
 $\tilde{\Phi }(p,t) = (2\pi )^{-1/2} \int\phi (y,t) \exp(-ipx/\hbar ) \, dy $, (iii) use an identity $\int_{-\infty }^{+\infty }
 \sqrt{a^2 + x^2}  \exp( \pm ipx) dx = - (2a/|p|) K_1(a|p|)$, \cite{brychkov}.

  The outcome presented in  Eq.~(2.24) of Ref.~\cite{remb},
  see also Eq.~(14) in Ref.~\cite{horwitz}, reads
$\sqrt{m^2c^4-\hbar^2c^2\frac{\partial^2}{\partial x^2}}\, \,  \phi (x,t) = \int _{-\infty }^{+\infty } K(x-y) \phi (y,t)\, dy$.
This result  is   plainly incompatible with  the primordial formulas (\ref{quasi}) and (\ref{quasi1}).  The same
  comment  (concerning the faulty outcome)  refers to an analogous reasoning  for  $\hbar c |\nabla |$, with the ultimate kernel
  $K(x-y) = -c\hbar /\pi (x-y)^2$.
 The crux is that the  above  arguments have  not been properly worked out.

Namely, while departing from the above mentioned identity (iii), let us evaluate an auxiliary integral:
\begin{eqnarray}
&&-\frac{1}{\pi}\int_{-\infty}^\infty \frac{K_1(a|p|)}{|p|}(e^{ ikp}-1)dp=\frac{1}{2\pi a}
 \int_{-\infty}^\infty dp\ \int_{-\infty}^\infty dx\ e^{ ipx}\sqrt{x^2+a^2}(e^{ ikp}-1)   =\nonumber \\
&&=\frac{1}{2\pi a} \int_{-\infty}^\infty \sqrt{x^2+a^2}\ dx \int_{-\infty}^\infty(e^{ ip(x+k)}-e^{ ipx})\ dp=
\frac 1a \int_{-\infty}^\infty [\delta(x+k)-\delta(x)]\sqrt{x^2+a^2}dx=\nonumber \\
&&=\sqrt{1+\left(\frac{k}{a}\right)^2}-1.\label{rfse6}
\end{eqnarray}
Since we  actually  have
\begin{equation}
\sqrt{m^2c^4-\hbar^2c^2\frac{\partial^2}{\partial x^2}}\  \phi(x,t) =\frac{mc^2}{\sqrt{2\pi}}\int_{-\infty}^\infty \tilde{\phi }(k,t) \ \sqrt{1+\frac{p^2}{m^2c^2}}\ e^{ikx}\, dk
\end{equation}
and an identification $a= mc/ \hbar $ implies $(k/a)^2= p^2/m^2c^2$, we realize that
\begin{eqnarray}
&&\sqrt{m^2c^4-\hbar^2c^2\frac{\partial^2}{\partial x^2}}\  \phi(x,t)=\frac{mc^2}{\sqrt{2\pi}}\int_{-\infty}^\infty dk\ \left(1-\frac{1}{\pi}
\int_{-\infty}^\infty \frac{K_1(\frac{mc}{\hbar}|y|)}{|y|}(e^{iky}-1)dy\right)e^{ikx} \tilde{\phi }(k,t)=\nonumber \\
&&=mc^2\left[\phi(x,t)-\frac{1}{\pi \sqrt{2\pi}}\int_{-\infty}^\infty  \frac{K_1(\frac{mc}{\hbar}|y|)}{|y|}dy
 \int_{-\infty}^\infty  dk\ (e^{ik(x+y)}-e^{ikx}) \tilde{\phi }(k,t) \right]=\nonumber \\
&&=mc^2\phi(x,t)-\frac{mc^2}{\pi} \int_{-\infty}^\infty  \frac{K_1(\frac{mc}{\hbar}|y|)}{|y|}[\phi(x+y,t)-\phi(x,t)] dy  . \label{rfse11}
\end{eqnarray}
This  implies a correct   integral  form of the 1D Salpeter equation  for $\psi (x,t) = \exp(imc^2t/\hbar ) \, \phi (x,t)$:
\begin{equation}\label{m}
i\hbar \frac{\partial \psi(x,t)}{\partial t}= [\sqrt{m^2c^4-\hbar^2c^2\frac{\partial^2}{\partial x^2}}  - mc^2] \  \psi(x,t)=
-\frac{mc^2}{\pi} \int_{-\infty}^\infty  \frac{K_1(\frac{mc}{\hbar}|y|)}{|y|}[\psi(x+y,t)-\psi(x,t)] dy
\end{equation}
Compare e.g.  Eqs.~(\ref{quasi}),   (\ref{quasi1}), see also   Eq.~(15)  in Ref. \cite{cufaro}.

Some comments are in order here. First, we  pay attention  that in   fact  we have accounted for
the  "missing  counterterm" from the very beginning of our calculations, by inserting
a factor $(e^{iky}-1)$ in the integrand.  One should realize that, if taken   literally, the integral
\[
I=\int_{-\infty}^\infty \frac{K_1(a|y|)}{|y|}dy
\]
is divergent,  even in the sense of the  Cauchy principal value.
 This is a consequence of  its $1/p^2$ behavior  of the integrand  as $p \to 0$, and its evenness.
On the other hand (set $a=mc/\hbar$) the  corresponding  "regularized" integrals (\ref{rfse6})  and (\ref{rfse11})  are  finite
\begin{equation}\label{rfse7}
-\frac{1}{\pi}\int_{-\infty}^\infty \frac{K_1\left(\frac{mc}{\hbar}|y|\right)}{|y|}(e^{ ipy/\hbar }-1)dy=\sqrt{1+\frac{p^2}{m^2c^2}}-1.
\end{equation}
For reference purposes, we list one more useful integral for   $\mu   \in (0,2)$, c.f. Eq. (\ref{intf2}):
\begin{equation}\label{rfse8}
|k|^\mu=\frac{1}{2\Gamma(-\mu)\cos\frac{\pi \mu}{2}}\int_{-\infty}^\infty\frac{dz}{|z|^{1+\mu}}(e^{ikz }-1).
\end{equation}
The Cauchy case is introduced through the limiting procedure  $\mu \rightarrow 1$, where
$\Gamma(-\mu)\cos\frac{\pi \mu}{2}$ $\to -\pi/2$ and  it follows
\begin{equation}\label{rfse9}
|k|=-\frac{1}{\pi}\int_{-\infty}^\infty\frac{(e^{ ikz}-1)\ dz}{|z|^2}.
\end{equation}
At this point we   realize   that   the   $m\rightarrow 0$  limit can be   executed  in (74). To this end we must  multiply
 \eqref{rfse7} by $mc^2$ and  take  notice  of   $K_1(x\to 0)=$ $1/x$.
Accounting for  $p=\hbar k$ ultimately reproduces  \eqref{rfse9}, next see Eq. (\ref{form}).
The corresponding Cauchy-Schr\"{o}dinger equation reads
\begin{equation}
i\partial _t \phi (x,t) = \hbar c |\nabla | \phi (x,t) = -\frac{\hbar c}{\pi} \int_{-\infty}^\infty  {\frac{[\phi(x+y,t)-\phi(x,t)]}{|y|^2}} dy  .
\end{equation}
 and provides a consistent definition of a spatially  nonlocal generator of quantum dynamics (Eq. (\ref{m}) likewise).
    Our discussion, including that of section IV.B,
 clearly demonstrates that the Fourier  transcription  of  spatially nonlocal expressions   (like e.g.  Eq. (\ref{intf3}))  is a derived  secondary
 ingredient of  the theory  which must not  be employed   hastily,   but with due care.

\subsection{Propagators  in 1D.}

L\'{e}vy semigroup kernels have the general form Eq.~(\ref{stable})  and (\ref{relativistic}.  The duality transformation $t\rightarrow it$ replaces
the semigroup operator by an affiliated unitary operator and likewise the Fourier representation of the  the semigroup kernel by  that of
 an affiliated propagator   (transition amplitude, unitary group kernel). Accordingly, keeping in mind that in general $k\equiv  k(x-y,t-s)$    with $t>s\geq 0$, we have:
\begin{equation}
k_{\mu }(x,t)  \rightarrow   k_{\mu }(x,it) = K_{\mu }(x,t)= (2\pi )^{-n}  \int \exp(-ipx - it|p|^{\mu })\, d^np \label{qstable}
\end{equation}
and
\begin{equation}
k_m(x,t)  \rightarrow  k_m(x,it) =  K_m(x,t) = (2\pi )^{-n}  \int \exp[-ipx - it (\sqrt{m^2 + p^2} - m)]\, d^np\, . \label{qrelativistic}
 \end{equation}
see e.g. \cite{wang} for a parallel discussion of the Gaussian (heat kernel vs free propagator) case.

First of all we must take under scrutiny, quite  often  met  in the literature,   appealing but  naive  short-cut   that amounts to  mimicking  the $t\rightarrow  it$ substitution of
Section  II.D  directly in the spatial expression for the  L\'{e}vy   semigroup kernel.  This procedure could have been justified in the Gaussian case   \cite{wang}  but is  invalid if
extended to   L\'{e}vy   kernels  without suitable precautions.

   The problem is that a formal analytic continuation in time
of  inverse polynomial pdfs     typically produces     singular   functions with  poles.
 Time honored quantum field theory procedures were developed in the past to
 handle similar   pole  problems, but  no mention  nor trace  of them could have been found  in  a  number of papers devoted to
  relativistic quantum mechanics,  \cite{horwitz,remb,cufaro,wiese}.

A fairly typical example   of   repeatedly reproduced erroneous outcomes are   e.g.   formulas  (4.31) and   (4.34) of Ref. \cite{wiese}
 and likewise (29), (C.5) and (C.19) in \cite{cufaro}. In particular, for  the   would-be    1D Cauchy  propagator,
  the   faulty  formula   $t/ \pi (x^2+t^2)  \rightarrow $
$ (1/2\pi )  \int \exp(\pm  ipx - it|p| )\, dp = it/ \pi (x^2-t^2) $ is  reproduced.
  We have addressed the involved Fourier integral before,
\cite{gar}. Presently, we shall    give  two complementary  derivations,  independent   from the previous one,  both   resolving   the poles problem.

We take the  previously mentioned  Cauchy kernel    $k_t=  t/ \pi (x^2+t^2)$  and rewrite it in the  form
\begin{equation}\label{ropp1}
k_t(x)=\frac{i}{2\pi}\left[\frac{1}{x+it}-\frac{1}{x-it}\right].
\end{equation}
To arrive at the Cauchy  propagator, we perform  a  formal substitution  $t \Rightarrow it+\varepsilon  $,
to be followed by the limit $ \varepsilon \to 0$.  This yields
\begin{equation}\label{ropp3}
K_t(\epsilon )=\frac{i}{2\pi}\left[\frac{1}{x- t+i\varepsilon}-\frac{1}{x+ t-i\varepsilon}\right].
\end{equation}
In view of the well known identity,  \cite{land},  ($P(1/x)$  indicates that the  generalized function  $1/x$  needs to be interpreted in terms of  the Cauchy principal value of the   involved  integral) ,
\begin{equation}\label{ropp4}
\frac{1}{x\pm i\varepsilon}=P\left(\frac 1x\right)\mp i\pi \delta(x),
\end{equation}
we get
\begin{eqnarray}\label{ropp5}
K_t (x)&=&\frac{i}{2\pi}\left[P\left(\frac{1}{x-t}\right)-i\pi \delta(x-t)-P\left(\frac{1}{x+t}\right)-i\pi \delta(x+s)\right]=\nonumber \\
&=&\frac{i}{2\pi}P\left[\frac{1}{x-s}-\frac{1}{x+s}\right]+\frac 12 \Biggl[\delta(x-s)+\delta(x+s)\Biggr]
=\frac{i}{\pi}P\left(\frac{t}{x^2-t^2}\right)+\frac 12 \Biggl[\delta(x-t)+\delta(x+t)\Biggr].
\end{eqnarray}
which is a correct expression for the Cauchy propagator, previously obtained in   \cite{gar}. That needs to be compared with the  naive
 outcome $it/ \pi (x^2-t^2)$, reproduced in \cite{cufaro,wiese}.
 We note that  the $\epsilon $-regularization still survives  in the  formula (\ref{ropp5}), although is not explicit. Its tacit
  presence is somewhat blurred
 by the Cauchy principal value indication  and the emergent  Dirac deltas.

Let us represent  the $\epsilon $-regularized Cauchy kernel  Eq. \eqref{ropp3} as   the following
Fourier integral
\begin{eqnarray}\label{ropp6}
K^{\epsilon }_t(x )&=&\frac{1}{2\pi}\left[\int_{-\infty}^0e^{i(x+t-i\varepsilon)u}du+\int_0^\infty e^{i(x-t+i\varepsilon)u}du\right]=
\frac{1}{2\pi}\int_0^\infty e^{-i(t-i\varepsilon)u}\left(e^{iux}+e^{-iux}\right)du\equiv\nonumber \\
&\equiv&\frac{1}{2\pi}\int_{-\infty}^\infty e^{-i(t-i\varepsilon)|u|}\cos ux \ du \equiv \frac{1}{2\pi}\int_{-\infty}^\infty e^{-i(t-i\varepsilon)|u|}e^{\pm iux} \ du  =  {\frac{i (t -i\epsilon )}{\pi  [x^2 -  ( t  -i \epsilon )^2]}}.
\end{eqnarray}

 It can be seen from \eqref{ropp6} that  the  $\epsilon $ - regularization is introduced  to secure he convergence of  integrals
  quantifying the   wave packet  evolution by means of kernel functions.  After  an explicit evaluation of the integrals, we can safely
   put $\epsilon \to 0$.  In view of this  implicit $\epsilon \rightarrow 0$ limit, the $-i\epsilon $ term in the numerator is in fact
   irrelevant and can be safely  neglected, yielding a familiar form of the regularized   1D quantum Cauchy  propagator
\begin{equation}
K^{\epsilon }_t \equiv  {\frac{ict }{\pi  [x^2 -  c^2( t  -i \epsilon )^2]}}\, .
\end{equation}
where we have reintroduced the velocity of light $c$-dependence.

 In the literature, one  encounters  an  appealing (in view of the  discussion of Section II.D) but  faulty  mapping
   of the Cauchy transition density into the Cauchy propagator $k_t=  t/ \pi (x^2+t^2)  \rightarrow  K_t = it/\pi (x^2-t^2)$,
   with no mention of the  pole problem and the need for a regularization,   see  for example  \cite{cufaro,wiese}.
    In fact,  if we  tacitly  disregard the pole  obstacle and, while  on the level of the Fourier representation,  formally  set
     $\varepsilon \to 0$ in  $K^{\epsilon }_t(x)$   and next   perform integrations in    Eq. (\ref{qstable})   (specialized to $n=\mu =1$),
     the outcome  would   actually   be    $K_t = it/\pi (x^2-t^2)$.
   As  indicated above, the fully-fledged   duality transformation  $\exp(-t|\nabla |) \rightarrow  \exp(-it|\nabla |)$
   enforces an $\epsilon $-regularization of the kernel function,  implying   less straightforward, but
    undoubtedly   correct outcome  Eq.~(\ref{ropp6}).

    The above reasoning extends to the quasirelativistic case as well.  Namely,   for the  pertinent
 semigroup  $\exp[-t(\sqrt{-c^2\Delta +m^2c^4} - mc^2)]$  the 1D kernel function (transition probability density of the associated jump-type process)
  reads:
 \begin{equation}
 k_m(x,t)= {\frac{mc^2t}{\pi }}
 \exp(mc^2t)\, {\frac{K_1(mc\sqrt{x^2 +c^2t^2})}{\sqrt{x^2+c^2t^2}}}
 \end{equation}
By turning over to the duality transformation $t\rightarrow it$, while  remembering that $K_1(z) \sim 1/z$ for small values of $z$,
 we can readily produce a regularized version of the quantum  propagator (note that we use $\equiv $ again)
 \begin{equation}
K^{\epsilon}_m(x,t)  \equiv  i {\frac{mc^2t}{\pi }}
 \exp(mc^2it)\, {\frac{K_1(mc\sqrt{x^2 -c^2(t -i\epsilon )^2})}{\sqrt{x^2  - c^2(t - i\epsilon )^2}}}\, .
 \end{equation}
whose $m\rightarrow 0$ limit coincides, as should  be the case,  with the previously defined quantum Cauchy propagator.
\begin{figure}
\centerline{\includegraphics[width=0.8\columnwidth]{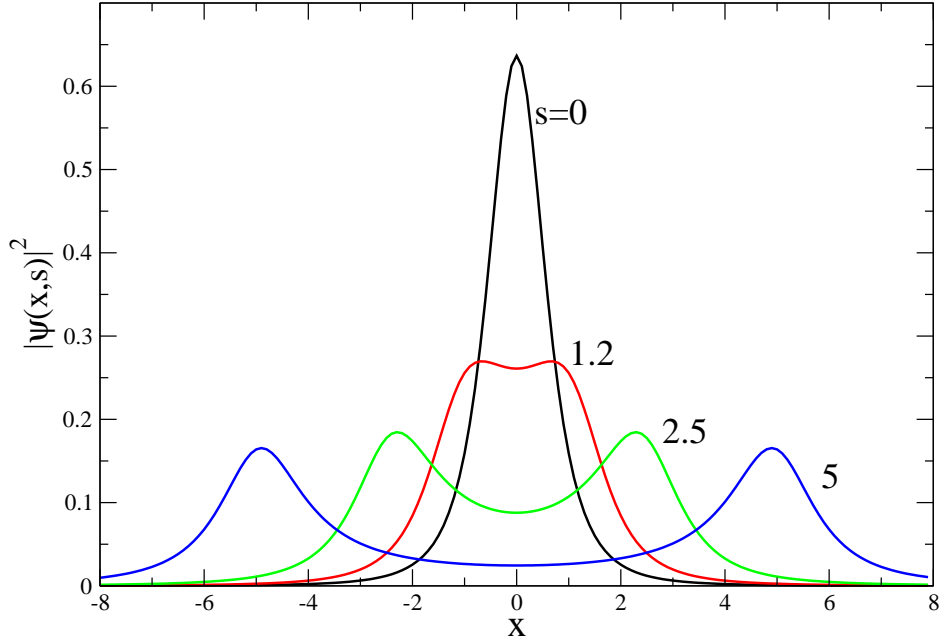}}
\caption{Time evolution of $\rho (x,t)=|\psi (x,t)|^2$, (\ref{ropp10}),  for the Cauchy-Schr\"{o}dinger dynamics (\ref{ini})
  with the  initial  data  (\ref{ropp7}).
Figures near curves show  time $t$ values.}
\end{figure}
 By turning back to the semigroup transport formulas (\ref{spectrum1}), (\ref{kernel}) and {\ref{conv}) and  directly executing
the duality mapping $t\rightarrow it$ right there, we can easily discover an origin of the  $\epsilon $-regularization, that is necessary if one
attempts  to deduce a consistent analytic expression for  quasirelativistic  and Cauchy  quantum  propagators.
 Let  us  focus on the Cauchy case again.  The unitary dynamics  of any $\psi (x,0)= f(x)$ is obtained as follows
 \begin{equation}
 \psi (x,t) = [\exp (-it|\nabla |) f](x) =(2\pi )^{-1/2}   \int_{-\infty }^{+\infty } \exp(-it|k| +ikx)\,
\tilde{f}(k) dk = (f*K_t)(x)
\end{equation}
where $ (f*K_t)(x): =\int  K_t(x-z)f(z)dz  $  and $K_t= (1/2\pi )\int \exp(-ikx - it|k|)\, dk $. As long as we do not insist on
an explicit evaluation of the Fourier integral expression for $K_t(x)$, no regularization is necessary.  The Fourier
expression for $\psi (x,t)$ evidently does the job without any special precautions.
On the other hand, would we   have  turned  to  $\psi(x,t) =(f*K_t)(x): =\int  K_t(x-z)f(z)dz  $,  it is only the regularized
expression (\ref{ropp5}) that yields a  time development of $\psi (x,t)$, in  conformity  with the Cauchy-Schr\"{o}dinger equation, \cite{gar}.
The  naive would-be propagator $K^{naive}_t (x)=it/\pi (x^2-t^2)$, if taken literally,  does not lead
 to  a consistent evolution pattern, see below.

\subsection{Quantum wave packet dynamics  in 1D }

  \subsubsection{Heavy-tailed initial data}
With the propagator in hands, we can address  the dynamical behavior of  solutions of the  free   fractional (eventually
Salpeter) Schr\"{o}dinger-type  equation. Lets us continue a discussion of  the   specific  1D  Cauchy-Schr\"{o}dinger  case
\begin{equation} \label{ini}
 i\partial _t \psi (x,t) =   |\Delta |^{1/2}  \psi (x,t)=  |\nabla | \psi (x,t)
\end{equation}
As an initial condition  we  take  an  "almost  Lorentz"   (up to normalization) distribution, so that $|\psi |^2(x,0)$   actually is
 a  normalized  quadratic Cauchy pdf, \cite{gar2}:
\begin{equation}\label{ropp7}
\psi(x,0)=f(x)=\sqrt{\frac{2}{\pi}}\frac{1}{1+x^2}.
\end{equation}
The integral kernel  of $\exp (-i |\nabla | t)$ takes    $\psi (x,0)$  into $\psi (x,t)$    through a convolution, c.f.  Eq.~ (\ref{conv}).
For clarity of discussion we  employ   a regularized form  $K_t^{\epsilon }(x)$ of the integral kernel, allowing to control its behavior in the vicinity of singularities,  so arriving at
\begin{equation}\label{ropp8}
\psi(x,t)=f(x)*K^{\epsilon }_t(x)\equiv \int_{-\infty}^\infty f(x-z)K^{\epsilon }_t(z)dz=\frac{1}{2\pi}\sqrt{\frac{2}{\pi}}
\int_{-\infty}^\infty \frac{dz}{1+(x-z)^2}\int_{-\infty}^\infty \ e^{-i(t-i\varepsilon)|u|}\cos uz \ du.\nonumber \\
\end{equation}
Consider   $\quad I=\int_{-\infty}^\infty \frac{\cos uz \ dz}{1+(x-z)^2}=
\pi e^{-|u|}\cos ux.\quad $  Thus, clearly
\begin{equation}
\psi(x,t)=\frac{1}{\sqrt{2\pi}}\int_{-\infty}^\infty e^{-|u|}\cos ux \ e^{-i(t-i\varepsilon)|u|}du \equiv
\frac{1}{\sqrt{2\pi}}\int_{-\infty}^\infty e^{-|u|}e^{iux} e^{-i(t-i\varepsilon)|u|}du.
\end{equation}
 and an explicit  evaluation   of   the   integral, followed by  $\varepsilon \to 0$,  gives rise to
\begin{equation}\label{ropp9}
\psi(x,t)= {\frac{1}{2}} [f(x+t)  + f(x-t)] + {\frac{i}{2}} [(x-t)f(x-t)- (x+t)f(x+t)] =  \sqrt{\frac{2}{\pi}}\frac{1+it}{(1+it)^2+x^2},
\end{equation}
where $f(x)$ stands for the initial data (\ref{ropp7}) for the pertinent evolution.  Taking   the modulus of a complex function \eqref{ropp9} we get
interesting outcome,  with an implicit redefinition   $\rho ^{1/2} _0(x)= f(x)$:
\begin{equation}\label{ropp10}
|\psi(x,t)|^2= (1+ t^2) \sqrt{\rho _0(x+s) \rho _0(x-s)} =\frac{2}{\pi}\frac{1+t^2}{\left[1+(x-t)^2\right]\left[1+(x+t)^2\right]},
\end{equation}
which conveys a useful information about an enhanced  spreading  of the initial wave packet   due to the dynamically generated  bimodality of the square root expression.
Compare e.g.  \cite{horwitz,gar,cufaro,remb}. The pertinent time-evolution is visualized in Fig.1, compare e.g. also  \cite{cufaro}.

We would like to point out  a definite advantage of the  ($\epsilon $)  regularized Fourier representation of the kernel which permits to
incorporate naturally the delta-function terms and  simplifies  calculations. A  complementary  detailed   derivation,
 explicitly  accounting for delta-type contributions  (\ref{ropp5}),
can  be found in Ref. \cite{gar}.

\subsubsection{Link with  a classical  (d'Alembert)    wave      equation. }

 The solution (\ref{ropp9}) of  Eq. (\ref{ini})   at the first glance  seems to  look like  a general solution of the  original
 equation of motion, represented  as a superposition  of  special solutions  (wave packets that are "moving" right or left  in the standard quantum lore),
   like e.g. $\Phi (x,t) = (1/\sqrt{2}[\phi (x-t) + \psi  (x+t)]$, c.f.  Eq. (3.37) in \cite{remb}.  This  is merely an illusion and we must consider
(\ref{ropp9}) as an {\it indivisible  whole},   since no  physical status  can be assigned    to  separate   right or left "moving"  components.
 This is a  direct   consequence of the manifest nonlocality of the motion generator $|\nabla |$.

 Our standpoint  is  that   there is no such entity as  the right or left  "moving particle"
 in the present   framework.     That stays  in plain opposition  with a discussion  of  Sections 3  and 4  in  Ref. \cite{remb}.
  Even worse, the very "particle"  notion (concept)    appears to be  doubtful  in this nonlocal dynamics    setting,   in view of the well known   "particle"
   (especially massless "particle")  localization problems plaguing the  relativistic quantum mechanics, c.f.  for example   Section V  of
   Ref. \cite{gar}and
   \cite{hawton,ibb} and references there-in.

  Eqs  (\ref{ropp9}) and (\ref{ropp10})  refer to  a nonlocal  complex-valued  wave phenomenon,  implying  that the induced   probability distribution
   bimodally    expands and   ultimately  fades away (spreads). The   underlying dynamical mechanism amounts to the    propagation of   local maxima,
  in  the  opposite directions,   with the velocity of light (here $c=1$), (\ref{ropp10}).
  None of those maxima  separately   can   be  given a status of physical relevance, e.g. {\it  must not   be} interpreted   as   an identifier of  a  right or  left
    "moving particle"  wave packet.

  On the other hand  the   left- and right "moving" components of  (\ref{ropp9})  can be given  a physical status within  the   higher level theory,  i. e.
the   pure wave 1D d'Alembert equation   $(-\Delta + {\frac{\partial ^2}{\partial t^2}}) u(x,t)=0$, where $c=1$.
 Both pertinent "moving" components are solutions of this wave equation and
 likewise   their superposition (\ref{ropp9})  is.   The crux   is that neither of  component functions is by itself a solution of the Cauchy-Schr\"{o}dinger
 equation.    It is  the "indivisible  whole",    i.e.  their superposition  $\psi (x,t) $  of Eq.~(\ref{ropp9}),   which  for sure   solves (\ref{ini})  and admits a    standard  quantum mechanical
 (Born's) probabilistic interpretation.

  It is worth pointing out that  the  Hilbertian   domain  of  the  operator  $|\nabla  |$  (containing all normalizable solutions  of $\partial _t\psi = |\nabla |\psi $)
  can be  consistently  built   by   putting through a suitable sieve the set  of   all   solutions of the 1D D'Alembert equation.    \
  The   pertinent  domain is  a  fairly  restrictively selected  subset  (closed linear space)    of  solutions of the   d'Alemebert equation,   such that    the Cauchy-Schr\"{o}dinger equation  is
    solved by them  as well (that we know  {\it  not necessarily} to be the case).

{\bf Remark:} With Eq. (\ref{ropp9}) in hands we can be more explicit on the last sentence of   the previous
subsection.
Namely, the Cauchy principal value of  the   (otherwise divergent)   integral representing a  "naive" propagation
 $\psi(x,t) =(f*K^{naive}_t)(x): =\int  K^{naive}_t(x-z)f(z)dz  $ of
$\psi(x,0)=f(x)=\sqrt{2/\pi}/(1+x^2)$   would result merely in the pure imaginary term  $(i/2) [(x-t)f(x-t)- (x+t)f(x+t)]$ of the expression (\ref{ropp9}),
which we know {\it  not to solve} the Cauchy-Schr\"{o}dinger equation as it stands, c.f. \cite{gar} for more details.

\subsubsection{Dynamically generated bimodality}

In view of Eqs.~(\ref{ropp9})  and (\ref{ropp10}),  a minor modification of the  initial data   (\ref{ropp7}) to    $\psi (x.0) =\sqrt{\frac{2\gamma }{\pi}}\, [\gamma / {\gamma ^2 +x^2}]$,  implies   $\psi (x,t)=    \sqrt{\frac{2\gamma }{\pi}} \left[ (\gamma +it)/ [(\gamma +it)^2 + x^2]\right]  $  and thence gives rise to
  the time-dependent pdf  of  the form
\begin{equation}
\rho (x,t) =   {\frac{2\gamma }{\pi}}  {\frac{  \gamma ^2 + t^2}{x^4 + 2x^2 (\gamma ^2 - t^2) + (\gamma ^2 + t^2)^2}} \, . \label{r}
\end{equation}
We  can quantify   an  emergence of    bimodality by investigating an extremum of $\rho (x,t)$  with respect to $x$,  i. e.  solving an equation $\partial _x  \rho = 0$. This amounts to solving $x^3 + x(\gamma ^2 - t^2) =0$ with an obvious outcome: $x_0=0$,
and two more real roots  defined by  $x_{\pm } = \pm \sqrt{ t^2- \gamma ^2}$ under the condition $t^2- \gamma ^2  >0$.

Accordingly, for times  $t^2< \gamma ^2 $ the  considered pdf  is unimodal,   while at time instants $ t^2 = \gamma ^2 =0$ the situation changes. The  bimodal form od the pdf is born and persists for all $t^2>\gamma ^2$.
The pdf of the previous subsection is recovered by putting $\gamma =1$.    We shall show subsequently that the reported behaviot of Cauhcy-Schr\"{o}dinger pdfs is not due to a special choice of initial data.
Additionally, we point out that the dynamical generation of bimodality is not special to the Cauchy-Schr\"{o}dinger case and appears as well in
 the quasirelativistic (Salpeter) evolution.\\

{\bf Remark 1:} There is no qualitative change in the behavior of solutions  if we pass to  the 1D  quasirelativistic  (Salpeter) equation  $\partial _t \psi (x,t) = \sqrt{-  \Delta  + m^2 } \psi (x,t)$.  Initial data of the form
\begin{equation}
 \psi (x,0) =   \left[ {\frac{m}{\pi K_1(2m\gamma )}}\right]^{1/2} {\frac{\gamma K_1 (m\sqrt{x^2+ \gamma ^2})}{\sqrt{x^2 + \gamma ^2}}}
\end{equation}
evolve in time according to a simple substitution rule $\gamma \rightarrow \gamma + it$.
As well, one easily verifies that the $m\rightarrow 0$ limit reproduces the Cauchy-Schr\"{o}dinger
 wave  function, whose $|\psi |^2(x,t)$ is indeed  the pdf  $\rho (x,t)$  of
Eq.~(\ref{r}), see e.g. \cite{usher,remb, cufaro}.
The denominator $\sqrt{x^2 +  (\gamma +it)^2}$    is a source of  an emergent    bimodality of this particular solution,  as  first noticed  in   Ref \cite{usher}.

{\bf Remark 2:} For both 1D Salpeter ($m >0$   and $m=0$)   cases,  a continuity equation   $\partial _t \rho = - \nabla j$   has been verified to make sense and suitable probability currents are known in a closed analytic form, see e.g. \cite{remb}.

\begin{figure}
\centerline{\includegraphics[width=0.73\columnwidth]{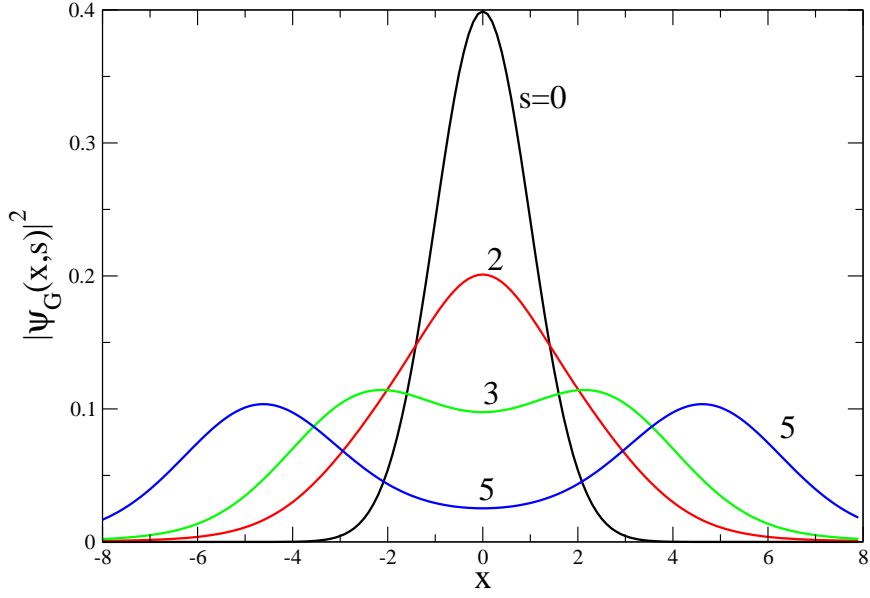}}
\caption{Time evolution of $\rho (x,t)=|\psi (x,t)|^2$  for the Cauchy-Schr\"{o}dinger dynamics (\ref{ropp13}),
  with the  initial  data  (\ref{ropp11}).
Figures near curves show  time $t$ values.}
\end{figure}

\subsubsection{Gaussian  initial data}
Now we consider the  Cauchy-Schr\"{o}dinger  time evolution of the initial Gaussian wave packet, c.f. \cite{babusci} for a  complementary discussion of this issue,
\begin{equation}\label{ropp11}
\psi(x,0)=\frac{1}{\sqrt[4]{2\pi}}e^{-x^2/4} \, .
\end{equation}
The dynamics is  made explicit   by  employing the Fourier representation of $\psi (x,t)$. Namely, we have .
\begin{eqnarray}
&&\psi (x,t)=\frac{1}{\sqrt{2\pi}}\int_{-\infty}^\infty \tilde{ \psi } (u,t)e^{iux}du\equiv \frac{1}{\sqrt{2\pi}}\sqrt[4]{\frac{2}{\pi}}\int_{-\infty}^\infty e^{-u^2} \ e^{-it|u|}\ e^{\pm iux}du \equiv \nonumber \\
&&=\left(\frac{2}{\pi}\right)^{3/4}\int_0^\infty \ e^{isz-z^2}\cos (zx) \ dz. \label{ropp12}
\end{eqnarray}
 The integral   can be evaluated exactly in terms of special functions to yield
\begin{equation}\label{ropp13}
\psi (x,t)=\sqrt[4]{\frac{2}{\pi}}\frac{e^{\frac{(x+t)^2}{4}}}{2\sqrt{2}}\left\{1+e^{tx}-
i\left[e^{tx}{\rm erfi} \left(\frac{t-x}{2}\right)+{\rm erfi} \left(\frac{t+x}{2}\right)\right]\right\},
\end{equation}
where ${\rm erfi} (x)=-i\ {\rm erf}(ix)$ is a real function.

We note in passing that following analytic methods of Ref. \cite{gar} we would obtain another explicit form of   the above $\psi (x,t)$:
\begin{equation}\label{ropp14}
\psi (x,t)=\frac{1}{2\sqrt[4]{2\pi}}\left\{\left[e^{-\frac{(x-t)^2}{4}}+e^{-\frac{(x+t)^2}{4}}\right] +\\
\frac{i}{2}P\left[\int_{-\infty}^\infty\frac{e^{-\frac{(x-y-t)^2}{4}}}{y}dy-
\frac{e^{-\frac{(x-y+t)^2}{4}}}{y}dy\right]\right\}.
\end{equation}

The  squared  modulus  (e.g. a corresponding pdf)  of the expression \eqref{ropp13} (or \eqref{ropp14}) is  displayed in in Fig.2.    A  qualitative similarity between Gaussian and Cauchy \eqref{ropp10} cases  is transparent.
 Both functions exhibit   a dynamically generated  bimodality, see   also \cite{cufaro}.   This  behavior of $\psi(x,t)$ is definitely  dictated by the propagator  and appears  not to depend on the initial data   $\psi(x,0)$  choice.

\begin{figure}
\centerline{\includegraphics[width=0.73\columnwidth]{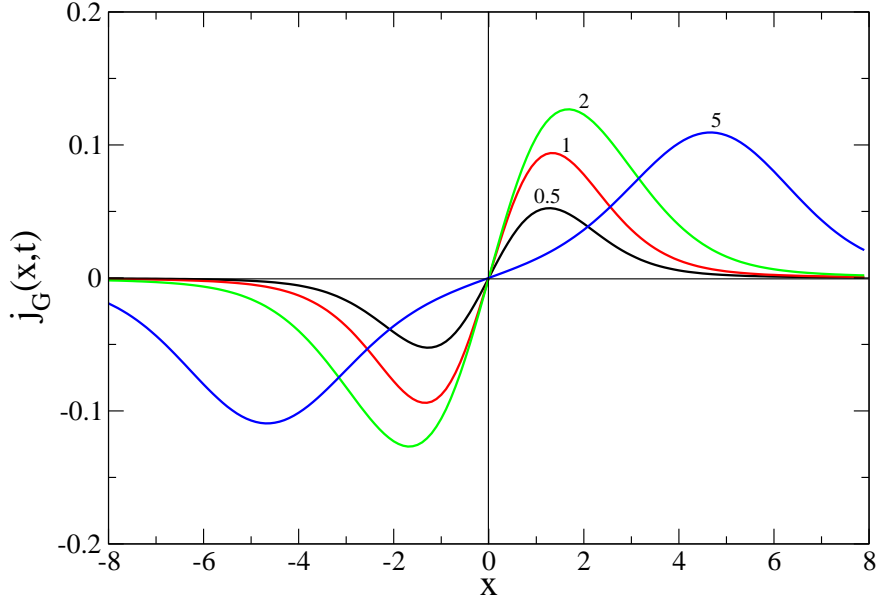}}
\caption{The probability current for an initially  Gaussian distribution in the 1D Cauchy-Schr\"{o}dinger dynamics.
Figures near curves show the time $t$ values.}
\end{figure}

We finally note that there is no problem to deduce,  with a numerical assistance if necessary,
  the  detailed time evolution  of any initial wave packet (Gaussian, Lorentzian etc) driven by
   a quasirelativistic kernel  (e.g. proportional to the MacDonald function $K_1$).

Let us calculate the probability current  for the present case (e.g. Gaussian initial data for the Cauchy-Schr\"{o}dinger dynamics.
To this end we  shall employ the observations  of \cite{remb}.
The Fourier transform of $\psi(x,0)=\frac{1}{\sqrt[4]{2\pi}}e^{-x^2/4}$ reads $\tilde{\psi }(p,0)=\sqrt[4]{\frac{2}{\pi}}e^{-p^2}$.
 The evolution can be encoded on the Fourier transform level as follows: $\tilde{\psi }(p,t) = \sqrt[4]{\frac{2}{\pi}}e^{-p^2-i|p|t}$.
Next, we have  to insert $\tilde{\psi }^*(p,t)$  and $ \tilde{\psi }(k,t)$ to the defining identity   (see e.g.  subsection   IV.E, here we scale away all dimensional constants  and consider 1D instead of 3D):
\begin{equation}\label{urk7}
j(x,t)={\frac{1}{2\pi}} \int_{-\infty}^\infty dp \int_{-\infty}^\infty dk \frac{p+k}{|p|+|k|}
e^{ix(k-p)}\tilde{\psi }^*(p,t)\tilde{\psi }(k,t).
\end{equation}
To facilitate the calculation of integral  we pass to the polar coordinates $p=r\cos\varphi$,
$k=r\sin\varphi$. Accordingly
\begin{equation}\label{urk8}
 j(x,t)=\frac{1}{\pi\sqrt{2\pi}}\int_0^{2\pi}\frac{\sin \varphi +\cos \varphi}{|\sin \varphi| +|\cos \varphi|}d\varphi
\int_0^\infty re^{-r^2}e^{ixr(\sin \varphi -\cos \varphi)}e^{itr(|\cos \varphi|-|\sin \varphi|)}dr.
\end{equation}

The integral in \eqref{urk8} can be evaluated analytically
\begin{equation}\label{urk9}
 J_r=\int_0^\infty r e^{-r^2}e^{i\lambda r}dr=\frac 12\left\{1-\frac{\lambda \sqrt{\pi}}{2}e^{-\frac{\lambda^2}{4}}
\left[-i+{\mathrm {erfi}}\left(\frac{\lambda}{2}\right)\right]\right\},\
\lambda(\varphi)=x(\sin \varphi -\cos \varphi)+t(|\cos \varphi|-|\sin \varphi|),
\end{equation}
so that the  final  expression for the probability current
assumes the form (cross-checked by means of  {\it Mathematica} routines):
\begin{equation}\label{urk10}
  j(x,t)=\frac{1}{(2\pi)^{3/2}}\int_0^{2\pi}\frac{\sin \varphi +\cos \varphi}{|\sin \varphi| +|\cos \varphi|}
\left\{1-\frac{\lambda (\varphi) \sqrt{\pi}}{2}e^{-\frac{\lambda(\varphi)^2}{4}}
\left[-i+{\mathrm {erfi}}\left(\frac{\lambda(\varphi)}{2}\right)\right]\right\}d\varphi.
\end{equation}
A numerical  visualization of the time development of that  current is reported in  Fig.~3.
 It is seen that behavior of the current is qualitatively similar
to that for the  Cauchy initial packet, see e.g.  Fig.2 in  Ref. \cite{remb}.
Namely, at small times,   there is practically no current at all (we begin from  $j(x,t=0)=0$). The current is
 an odd function of $x$. As time passes,  two positive and negative current peaks  develop and
move towards plus and minus infinities respectively,  while falling  down to zero.

\subsection{Quantum wave packet dynamics   in 3D.}

\subsubsection{Generic wave packet evolution  pattern: radial expansion.}

  Peculiarities of the   3D dynamical behavior can be conspicuously seen in the Cauchy dynamics $i\partial _t \psi = |\nabla | \psi $, where
   due to the radial symmetry, the   1D bimodality is replaced by a "spherical"-modality
   (and "circular"-modality  in 2D). The   initial   wave packet   rapidly  delocalizes
     by  a dynamically developed  expansion  of a   spherically-shaped probability concentration
       to the radial infinity, mimicking an expanding spherical wave with the source at the origin.
         Its  radial  peak is   running  away   with  the velocity $c$ (here
    interpreted as $c=1$):  the pdf  radial    maximum resides on a  surface of an
     expanding sphere (an expanding circle in 2D and an  expanding interval in 1D).

        This   property  can be directly deduced  from  the analytic formula  ($r=\sqrt{x^2+y^2+z^2}$)
    \begin{equation}\label{data}
\psi (r,0) = {\frac{(\sqrt{2\gamma })^3}{\pi }} {\frac{\gamma }{(r^2 + \gamma ^2)^2}}\rightarrow \psi (r,t)) =
 {\frac{(\sqrt{2\gamma })^3}{\pi }} {\frac{\gamma +it }{[r^2 + (\gamma  +it)^2]^2}}
\end{equation}
implying (set $\gamma =1$) the following functional form of the related pdf:
 \begin{equation}
\rho (r,0) = |\psi (r,0)|^2   ={\frac{8}{\pi ^2}} {\frac{1}{(1 + r^2 )^4}}  \longrightarrow                         \rho (r,t) = |\psi (r,t)|^2 = {\frac{8}{\pi ^2}} {\frac{1+t^2}{[(r^2-t^2+1)^2 + 4t^2]^2}}
  \end{equation}
Its  time evolution  is depicted in Fig. 4. ($\gamma =1$,  $r=\sqrt{x^2 +y^2+z^2}$).

\begin{figure}
\centerline{\includegraphics[width=0.73\columnwidth]{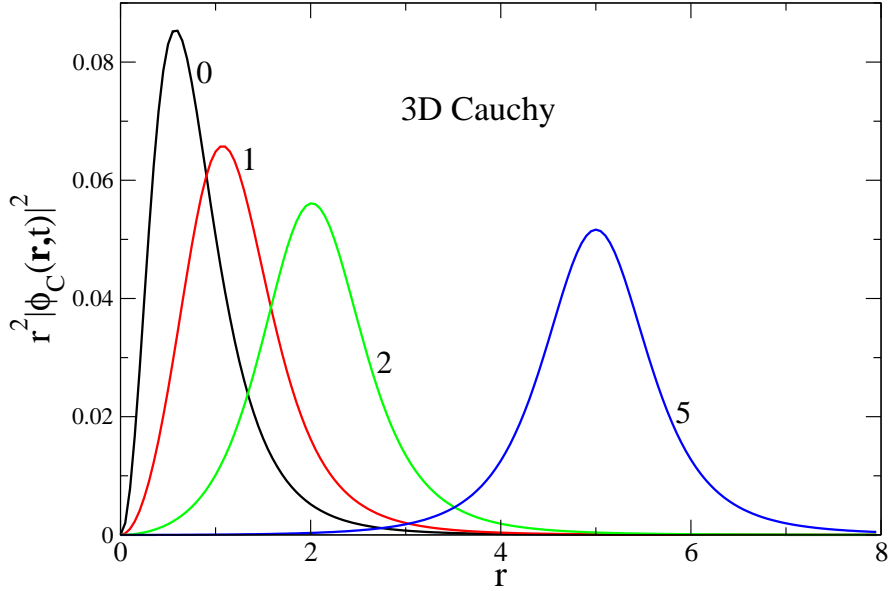}}
\caption{3D Cauchy-Schr\"{o}dinger evolution:  spherical expansion, exemplified for the radial pdf $r^2\rho (r,t)$}
\end{figure}

  We note that the normalization coefficient reflects the radial symmetry of the problem
  ($4\pi $ factor is the remnant of angular integrals), and Fig. 4  actually  depicts the behavior of $r^2 \rho(r,t)$,
   instead of the "plain"$\rho (r,t)$.  That should be compared with the time evolution pattern reported in Fig.~9 of \cite{remb}.

   The probability current for this case has received a closed analytic
   form in \cite{remb}.  Both the dynamical behavior and the rapid decay of $j(x,t)$ are depicted in Fig.~10 of \cite{remb}.

{\bf Remark:}  Since a solution of the Cauchy-Schr\"{o}dinger equation is also a solution of the D'Alembert equation (and not necessarily
 in reverse, c.f. subsection  IV.E.2),  it is worth pointing out that the pertinent wave equation is known  to admit radially   expanding (or  contracting)
  spherical   waves that  originate  from (or sink in)  a point source (or sink).
   Superpositions of   expanding and contracting waves are legitimate solutions as well.

\subsubsection{Propagators}

In analogy to considerations of  subsections  III.C and  IV.D,   we may  implement the  quantum  propagation    $\psi (r,0) \rightarrow \psi (r,t)$
 by means of the 3D propagator.  To this end let us recall that  the Cauchy kernel   (transition probability density of the Cauchy jump-type process)
  stems form the general formula (\ref{cauchy}), specialized to $n=3$.   Accordingly,  while making  explicit the $c$-dependence,    we depart from
\begin{equation} \label{1}
k_0({\bf x }- {\bf x'}, t-t') = {\frac{ c(t- t')}{  \pi ^2 [({\bf x} - {\bf  x'}) + c^2(t-t')^2]^2}}= {\frac{1}{(2\pi )^3}}
\int d^3k  \exp [-i{\bf k}({\bf x}  - {\bf x'})  -  |{\bf k}|   c(t - t')]
\end{equation}
where $|{\bf k}| =  \sqrt{k_x^2 +k_y^2+ k_z^2}$.  For clarity of further discussion we shall keep in mind a space-time homogeneity of the corresponding random  motion and  retain ${\bf x}$ and $t$  labels, instead of ${\bf x}  - {\bf x'})$ and $t-t'$ respectively.

  The duality transformation   $\exp(- c|\nabla  |t)  \rightarrow  \exp(-ic |\nabla |t)$ can be accomplished on the    level of integral kernels, provided we take care of potentially dangerous singularities (poles). Like before that  can be secured
  by considering an $\epsilon $-regularized   Cauchy-Schr\"{o}dinger propagator   in the form (compare e.g. also \cite{garcz}):
  \begin{equation}
  K^{\epsilon }_t ({\bf x}) =  {\frac{1}{(2\pi )^3}}   \label{2}
\int d^3k  \exp [-i{\bf k} {\bf x}   -  ic  |{\bf k}|( t  -i\epsilon) ]  \equiv  {\frac{ict}{\pi ^2 [r^2 - c^2(t-i\epsilon )]^2}}\, , \end{equation}
where   $r= |{\bf x}|$.  C. f.   also   our comment following Eq.~(\ref{ropp6}) in regard to  $\equiv $  versus  $=$   in the above formula.

Perhaps it is instructive to mention (and make a direct comparison with Eq. (\ref{2})  the   so called complexified fundamental solution
  of the d'Alembert equation,
\cite{ibb1}, which has  the form (we set $\epsilon $ instead  of the  originally employed  real constant $a$)
\begin{equation}
K_{\epsilon }({\bf x},t)  = {\frac{1}{c^2(t- i \epsilon )^2  - r^2}}\, .
\end{equation}
For $\epsilon \neq 0$ this function is analytic in the whole space-time. Would we proceed formally (it is  not quite legal step in view of our
discussion of the poles issue), and set $\epsilon =0$, one arrives at the Hadamard fundamental solution which is singular on the light cone.

Armed with the  lesson   of subsection III.D and the above   (\ref{1})  to  (\ref{2})   realization of the  duality mapping, we know how to handle  the arising singularity problems in the 3D   ($m>0$)   Salpeter case as well.   Namely, the quasirelativistic jump-type process transition kernel  (e.g. probability density, whose  space-time  homogeneity is implicit),  in view of    Eq.~(\ref{k}) reads
\begin{equation}
k_m({\bf x}, t) =  {\frac{1}{(2\pi )^3}}
\int d^3k  \exp [-i{\bf k}{\bf x}  - t (\sqrt{c^2{\bf k}^2 + m^2c^4}- mc^2 )]     =   t\, {\frac{m^2c^3}{2\pi ^2}}\exp (mc^2t) \, {\frac{K_2(mc\sqrt{r^2 + c^2t})}{r^2   + c^2t^2}}
\end{equation}
where $K_2(z)\rightarrow 1/z^2$ for small values of $z$.

Accordingly, the $\epsilon $- regularized propagator of   the  3D   Salpeter   equation   $\partial _t\psi =  (\sqrt{-c^2\Delta + mc^2} - mc^2)\psi  $    reads:
\begin{equation}
K^{\epsilon}_m({\bf x}, t)  \equiv    i t\, {\frac{m^2c^3}{2\pi ^2}}\, \exp (mc^2 it) \,
{\frac{K_2(mc\sqrt{r^2   - c^2(t-i\epsilon )})}{x^2   - c^2(t-i\epsilon )^2}}
\end{equation}
where $r= \sqrt{x^2 + y^2 +z^2}$ and $\hbar=1$ has been presumed.  One  may  readily verify the validity of the $m\rightarrow 0$ limit, Eq, ~(\ref{2}).

\subsubsection{Probability currents in 3D}
For the  completeness  of exposition we find instructive to present a detailed derivation of the probability current  for the
 3D Salpeter ($m\geq 0$) equation, whose main idea we  borrow  from Ref. \cite{remb}.
 The transition to  a  simpler 1D case is obvious.    We represent the Salpeter dynamics  with  an additive perturbation by an   external potential
 \begin{equation} \label{ov1}
i\hbar \frac{\partial \psi}{\partial t}=\left[\sqrt{m^2c^4-\hbar^2c^2\Delta}+V({\bf x})\right]\psi
\end{equation}
Accounting for
\begin{equation}\label{ov2}
\psi^* \frac{\partial \psi}{\partial t}=-\frac{i}{\hbar}\left[\psi^*\sqrt{m^2c^4-\hbar^2c^2\Delta} \ \psi+V({\bf x})\psi^*\psi\right].
\end{equation}
and its  complex conjugate   we  can evaluate the  time derivative of the probability density
 $ \rho =  |\psi|^2$   according to
$ \frac{\partial(\phi^* \psi)}{\partial t}=\psi \frac{\partial \psi^*}{\partial t}+\psi^*
\frac{\partial \psi}{\partial t}$  so that
\begin{equation}\label{ov6}
-  \frac{\partial |\psi|^2}{\partial t}
=\frac{i}{\hbar}\left[\psi^*\sqrt{m^2c^4-\hbar^2c^2\Delta}\, \psi-\psi \sqrt{m^2c^4-\hbar^2c^2\Delta}\,
\psi^*\right].
\end{equation}

Our aim is   to transform    (\ref{ov6})  into a regular  3D continuity equation    $-\frac{\partial \rho}{\partial t}$$=\nabla {\bf j}$, $\rho=|\psi|^2$.
Taking the Fourier transforms  of each  $\psi $- entry  in   the right-hand-side of  (\ref{ov6}) separately,  and presuming that the r-h-s   actually  stands   for $ \nabla {\bf j}$,
 we  get
\begin{eqnarray}
&&\nabla {\bf j} ={\frac{ic}{(2 \pi)^3 \hbar^7}} \int d^3p\,  \, d^3k \, [\phi ^*_p\sqrt{m^2c^2 + k^2}\,  \phi _k - \phi _k \sqrt{m^2c^2 +p^2}\,  \phi _p^*]   \,
e^{i {\frac{{\bf x}}{\hbar}({\bf k}-{\bf p}}})=\\
&&
 {\frac{ic}{(2 \pi)^3 \hbar^7}}  \int d^3p\,  \, d^3k \,  \phi^*_p \phi_k \left[
{\frac{k^2-p^2}{\sqrt{m^2c^2+k^2}+\sqrt{m^2c^2+p^2}}} \right]    e^{i {\frac{{\bf x}}{\hbar}({\bf k}-{\bf p}}}) \, . \label{ov8a}
\end{eqnarray}
In view of $({\bf k}+{\bf p})({\bf k}-{\bf p})= k^2 - p^2$, we  readily retrieve the functional  (spectral)  expression  for   $\bf j$  that obeys the continuity  equation:
\begin{equation} \label{ov10}
{\bf j}=\frac{c}{(2 \pi)^3 \hbar^6}\int d^3p \ d^3k \frac{{\bf k}+{\bf p}}{\sqrt{m^2c^2+k^2}+\sqrt{m^2c^2+p^2}}\phi^*_p \phi_k e^{i\frac{{\bf x}}{\hbar}({\bf k}-{\bf p})}
\end{equation}
 The  massless case  arises if we set  $m=0$ in \eqref{ov10}. The  probability current for the massless case reads
\begin{equation} \label{ov11}
{\bf j}=\frac{c}{(2 \pi)^3 \hbar^6}\int d^3p \ d^3k \frac{{\bf k}+{\bf p}}{|{\bf k}|+|{\bf p}|}\phi^*_p \phi_k e^{i\frac{{\bf x}}{\hbar}({\bf k}-{\bf p})}.
\end{equation}
To have a   direct comparison with   some of our  previous  discussions, one should set $\hbar = 1 = c$ whenever necessary.

{\bf Remark:} In the present subsection we have discussed  the quasirelativistic generator that is additively perturbed by a function
 (like e.g.   a Newtonian potential) that secures a bounded-ness of the Hamiltonian in (\ref{ov1}) from below and
 (via an additive renormalization) moves its spectrum to $R_+$, with $0$ as its lowest eigenvalue.
  A number of typical (e.g. simplest) spectral  problems have been addressed in the past both for the
   3D and 1D quasirelativistic cases, \cite{remb1,lucha,lucha1,lucha2,kaleta} and references there-in.
Some special spectral problems  (harmonic potential and the infinite well case) for
the $m=0$,  were considered e.g. in
 \cite{laskin,laskin1,remb1,gar1,gar2,malecki} and \cite{jeng,baym,hawkins}. There are open problems even in those simple
 cases; controversies concerning  an analytic form  of eigenfunctions and the ground state function in particular have not been resolved for a
 "particle in a box", \cite{laskin,baym,hawkins}. Mathematical arguments \cite{kwasnicki}, indicate that none of proposed so far
  solutions   is correct,  as confirmed in the  very recent Ref. \cite{luchko}.

\subsubsection{Non-existence of the continuity equation for  the fractional  quantum dynamics.}

In the standard quantum mechanics, where the minus Laplacian  is   recognized  to serve also   as  the Wiener noise generator, the time derivative
of $\rho $ is routinely represented in terms of the continuity equation.  The argument goes as follows: consider a normalized solution
 of the Schr\"{o}dinger equation  (we set here-by $\hbar/2m = 1$), then $ \partial _t |\psi |^2 = i [\psi ^*(\nabla ^2)\psi  - \psi \nabla ^2 \psi ^*]=
 i \nabla (\psi ^*\nabla \psi - \psi \nabla \psi ^*)= - \nabla j$.

 If we consider  the $\mu $-stable    generator instead of the minus Laplacian,  a   consequence of the L\'{e}vy-Schr\"{o}dinger  equation is:  $\partial _t |\psi |^2 = - i(\psi ^*|\Delta |^{\mu /2}\psi  - \psi |\Delta |^{\mu /2} \psi ^*)$.
 This equation readily applies to  the  $|\nabla | =  |\Delta |^{1 /2}$  and extends to the Salpeter generator $\sqrt{- \Delta  + m^2 }$, whose mass $m=0$   version  the  Cauchy generator actually is.
  The  just derived 3D version of the probability current can be associated with those  two
   special  cases.

  The situation appears more troublesome  when   general   L\'{e}vy-stable generators $|\Delta |^{\mu /2}$  are involved.
We note that in this case there holds (we refer to 3D, dimensional constants are scaled away)
\begin{equation}\label{j}
-  \frac{\partial |\psi|^2}{\partial t}
= \frac{i}{\hbar}\left[\psi^*|\Delta |^{\mu /2}\, \psi-\psi | \Delta |^{\mu /2}\, \psi^*\right] = {\frac{i}{(2 \pi)^3 }} \int d^3p\,  \, d^3k \,   {\frac{ |k|^{2\mu }     -   |p|^{2\mu }}{ |k|^{\mu }     +  |p|^{\mu }} }\, \phi ^*_p \,  \phi _k
e^{i {\frac{{\bf x}}{\hbar}({\bf k}-{\bf p}}}).
\end{equation}
and there is no way to extract  the  crucial  factor    ${\bf k}-{\bf p}$,   needed to identify (\ref{j})
 as a  legitimate (spectral)  expression for $\nabla {\bf j}$ , with a notable exception of $\mu =1$.

A candidate expression for  the  probability current,  albeit   restricted to the stability parameter range $\mu \in (1,2)$,    has been   produced in Ref. (\cite{laskin,laskin1}, see also an  explicitly  uncompleted   passage from (38) to (39) in  \cite{dong}.
 While setting $\hbar \equiv 1$ and disregarding the noise intensity parameter  $D_{\mu }$,  $ {\bf j}$
 has been presented in the form:
\begin{equation}
{\bf j} = -i  \psi ^*  |\Delta |^{\mu /2 -1}  \nabla   \psi   - \psi    |\Delta |^{\mu /2 -1}   \nabla   \psi ^*     \label{jl}
\end{equation}
The problem   is  that the   identity  $\partial _t \rho = - \nabla {\bf j}$  is invalid  as it stands, in its differential form, e.g. {\it there is no}   $-div {\bf j}$  local   representation of $\partial _t\rho $.
  This  can be   demonstrated by    inspection   after passing to the  spectral representation.

Our negative argument goes as follows, \cite{zaba}.  First we observe that by taking the divergence   of ${\bf j}$, (\ref{jl}), we  get
\begin{equation}  \label{gradj}
\nabla {\bf j} = -i \left[ (\nabla \psi ^*) \, |\Delta |^{\mu /2 -1}  \nabla   \psi  -  (\nabla \psi ) \,  |\Delta |^{\mu /2 -1}   \nabla   \psi ^* \right]    -  i \left[ \psi ^*\, \nabla |\Delta |^{\mu /2 -1}  \nabla   \psi   - \psi \, \nabla  |\Delta |^{\mu /2 -1}   \nabla   \psi ^* \right]
\end{equation}
 Now,   we shall  consider  the second square-bracketed term  in the above. To this end   let us  explicitly evaluate the action of involved operators, while passing to  the spectral (Fourier) representation of the relevant entry.
We have e.g.
\begin{equation}
\psi ^*\, \nabla |\Delta |^{\mu /2 -1}  \nabla   \psi  = -   \psi ^*({\bf x})   {\frac{1}{(2\pi )^{3/2}}}  \int d^3p  \,  |p|^{\mu } e^{i{\bf p x}} \, \tilde{\psi }({\bf p})   =  -   \psi ^*|\Delta |^{\mu /2}\psi    \label{f}
\end{equation}
and taking into account a complex conjugate of    (\ref{f}),  we recover the right-hand-side  of the   transport  equation  (\ref{j}), (up to $\hbar \equiv 1$  and an obvious sign inversion).  For ${\bf j}$ of Eq. (\ref{jl}) to be a
legitimate probability current we must  guarantee  that the first square bracketed term of Eq.  (\ref{gradj})  identically vanishes. This is not the case for   complex functions $\psi $.

{\bf Remark:}  It has been demonstrated in \cite{gar}, c.f. also \cite{angelis}, that   the  transport equation
 $\partial _t |\psi|^2  = - \frac{i}{\hbar}\left[\psi^*|\Delta |^{\mu /2}\, \psi-\psi | \Delta |^{\mu /2}\, \psi^*\right]$
  can be literally  represented as a master-type equation for the
jump-type process with a corresponding  jumping  rate  $q({\bf x},t,A)$ where $A$ is a Borel set in $R^3$.
The  pertinent  master-type equation has the form  $\partial _t \rho (A,t) = \int q({\bf y},t,A) \, \rho ({\bf y},t) \, d^3y$  and $q({\bf x},t,A)$
critically depends on the  involved  L\'{e}vy measure.
If $\rho ({\bf x},t)$ is a probability
density of a certain stochastic  process, then $\int_A\rho(x,t)d^3x= \rho (A,t)$ tells us what is the probability for a jump to have its spatial direction and size matching
${\bf x} \in A $.
However, one should not be inclined to think that  a concrete jump refers to a "physical particle" that jumps in space.  We shall leave this debatable point
 open for further discussion.  To the contrary, in a conventional
Laplacian-based  quantum mechanics,  a standard interpretation of $\rho ({\bf x},t) (\Delta x)^3$  as a probability  to locate
a "physical particle" in a  cube of volume  $(\Delta x)^3$ seems to be consistent.

\section{Domains for nonlocal operators, state vectors   and  uncertainty relations}

\subsection{Mathematical prerequisites}

Hamiltonian  operators discussed
in this paper are unbounded. When defining an unbounded operator,
it always is necessary to specify its domain of definition.
We proceed in he following general manner.  If $A$
is an operator in the Hilbert space ${\cal{H}}$, we write $D(A)
\subset {\cal{H}}$ for the domain of $A$.

Let us consider a densely defined  self-adjoint
operator $A$. For any $g\in D(A)$, we set
\begin{equation}\label{scalar}
||g||_A = [(Ag,Ag) +
(g,g)]^{1/2}\, .
\end{equation}
 Then $||\cdot ||_A$  is a norm in $D(A)$.   As  well, we consider  $(f,g)_A=
(Af,Ag) + (f,g)$ to be a scalar product specific to  $D(A)$.

If $f_n
\in D(A)$ is a Cauchy sequence in $||\cdot ||_A$, that is,
$\lim_{n,m \rightarrow \infty} ||f_m - f_n||_A =0$, then $f_n$
also is a Cauchy sequence in the Hilbert space ${\cal{H}}$ norm
$||f|| = [(f,f)]^{1/2}$. By the completeness of ${\cal{H}}$ there
is $f\in {\cal{H}}$ such that $\lim_{n\rightarrow \infty}
||f-f_n|| = 0$. It follows that $f$ necessarily belongs to
$D(A)$ (that is, $D(A)$ is complete in the $||\cdot ||_A$ norm).  Then
 $A=A^*$ is closed and denoted  $A = \overline{A}$.  (The closedness is a consequence of
  $D(A)= D(A^*)$ and an assumption that  D(A) is dense in ${\cal{H}}$, \cite{karw1}.)

In  our text, we  have  introduced   Hamiltonian-type  operators $\hat{H} = \hat{H}_0  + V$,
 where   $\hat{H}_0$  is a nonlocal generator of L\'{e}vy noise,  via a canonical quantization
 of the L\'{e}vy-Khintchine formula. Consequently,  we encounter standard quantum mechanical domain
  problems for unbounded operators. This refers  as well  to  $\hat{H}$ and
 position-momentum (fairly standard !) operators   $\hat{x}, \hat{p} = - i\nabla $ which form
  a canonical pair $[\hat{x}, \hat{p}] \subset iI$,  up the presence of $\hbar$ which in  various parts of the paper has been identified with unity..

Let us take  Hilbert space vectors $f, g \in L^2$ and let  $(f,g)\in C$   represent   a standard  scalar
product in the complex  functions space.
 We   need
 \begin{equation}
 (f,\hat{H}g)= (f,\hat{H}_0g) + (f,Vg)
 \end{equation}
  to have granted the existence  status (e. g. the   convergence of involved integrals).
In particular, if  $f$ is  a normalized function,  we   may  expect the mean energy  value $(f, \hat{H}f)= \langle H_0\rangle + \langle V\rangle $
here-by  defined  up to dimensional constants, to exist.
Since our main objective are domain issues for $\hat{H}_0$, let us set $V=0$ and pay attention to $(f,\hat{ H}_0 g )$  exclusively.

In passing  we note that  in reference to  the semigroup  dynamics  and the associated jump-type processes,
  a well developed  mathematical formalism of  Dirichlet forms associated with  noise generators $-\hat{H}_0$ is available, \cite{lorinczi,liming,chen}.
   The pertinent quadratic forms "live" in a Hilbert space of real-valued functions.
  This subject matter has been  also addressed in the physics-oriented literature but exclusively in
  relation to  $- \Delta $,  i. e. in
   the context of   the Wiener noise and  the   emergent  semigroup description of the   Brownian motion, \cite{albeverio}.

  We are somewhat guided by basic tenets  of that theory, but our considerations will refer directly to the unitary  (quantum) dynamics  and  therefore our Hilbert space contains complex valued $L^2$ functions, the real ones
   forming  merely  a subset  in $L^2$  and  {\it  not } a linear subspace, \cite{karw}.

   All  symmetric   L\'{e}vy   noise generators are  Hermitian  (actually self-adjoint, \cite{applebaum}),   $(f, \hat{H}_0g) =  (\hat{H}_0f,g)$,
   on their  domains od definition. They are also  non-negative. Therefore their operator  square roots
    are always well defined as  self-adjoint operators.

   Accordingly, we can pose an auxiliary domain problem  for $H_0^{1/2}$,  by considering
   \begin{equation}\label{square}
(f,\hat{H}_0g) = (\hat{H}_0^{1/2}f,\hat{ H}_0^{1/2}g)
\end{equation}
as a relevant (e.g. non-trivial)   part of the
    induced  scalar product  $(f,g)_{H_0^{1/2}}$  in   $ D(H_0^{1/2}) \subset L^2$,
     c.f. Eq.~(\ref{scalar}).

     The operator domain of interest would consist of all $f$ and $g$ in  $ L^2$, such that
      the completion of a corresponding linear space in the $\|  . \| _{\hat{H}^{1/2}_0}$  norm would
        identify a closed  dense subspace of $ L^2$.
         We shall  not be more elaborate on purely mathematical issues and assume such
          domain existence for granted, for each  self-adjoint  $\hat{H}_0$  of  Section III separately.

\subsection{Test model: Quasirelativistic Hamiltonian.}

 As a test model let us consider  the quasirelativistic Hamiltonian $H_m= \sqrt{m^2 - \Delta  } -m$,
  ($\hbar = c = 1$).   We know  that solutions  of the Salpeter equation
   \begin{equation}\label{m}
   i\hbar \frac{\partial \psi(\mathbf{x},t)}{\partial t}= [\sqrt{m^2- \Delta }  - m] \  \psi(\mathbf{x},t)
   \end{equation}
  have the form $\psi (\mathbf{x},t) = \exp(imt ) \, \phi (\mathbf{x},t)$.
   Therefore,  instead of  introducing the domain for $H_m$ proper, which is  a nonnegative operator
    (with $0$ as the bottom  generalized  eigenvalue), we may   pass to an  equivalent   domain analysis,
     carried out for the strictly positive operator  $H_m+m=  \sqrt{m^2 - \Delta }$,
   see e.g. our discussion of Section IV.B.2.

  This step is advantageous, since if  $\psi (\mathbf{x},t)$ is a solution of the
  pseudodifferential--Schr\"{o}dinger
  equation  $i\partial _t\psi = [\sqrt {-\Delta + m^2} -m]\psi $, then
   $\phi (\mathbf{x},t) = \psi (\mathbf{x},t)\exp(-imt)$    necessarily is a
   positive energy solution of the free Klein-Gordon equation $(\Box +m^2)\phi
  (\mathbf{x},t)=0$, with $\Box =  - \Delta + \partial ^2/\partial t^2 $.

  Each \it scalar \rm positive energy solution $\phi (\mathbf{x},t)$ of the free Klein-Gordon
  equation can be represented in the manifestly Lorentz covariant form:
  \begin{equation} \label{klein}
  {\phi (\mathbf{x},t)={1\over {(2\pi )^{3/2}}} \int d^4k\,
  e^{(-ik_{\mu }x^{\mu })}\, \delta (k_{\mu }k^{\mu } - m^2) \Theta (k_0)\,
  \Phi (k_0,\mathbf{k})} ={\frac{1}{(2\pi )^3}} \int {\frac{d^3k}{k_0}}
  \Phi (k_0,\mathbf{k}) e^{(-itk_0 + i \mathbf{k} \mathbf{x})}
  \end{equation}
 where $k:=(k_0,\mathbf{k})$, $k_{\mu }k^{\mu }:=k_0^2 - \mathbf{k}^2$, $\Phi (k)$
 is a scalar and
 $\Theta (k_0)$ is the Heaviside function equal to $1$ if $k_0>0$ and
 to $0$ otherwise. Ultimately we are left with   $k_0= \sqrt{\mathbf{k}^2 + m^2}$.

 This  representation extends to all solutions of $i\partial _t \phi = \sqrt {-\Delta +m^2} \phi $,
 and upon changing $k_0\to -k_0$ in $\Theta (k_0)$ followed by a complex conjugation, to solutions
 of the time adjoint equation as well. (Side comment:  general solutions of those
 pseudodifferential--Schr\"{o}dinger
 equations form Lorentz invariant subspaces in the linear space  of all solutions to the free
 Klein-Gordon equation.)

  Let us tentatively  adopt the following definition of the Klein-Gordon scalar
 product, \cite{barut}, (being  independent of the
 specific space-like surface of integration):
 \begin{equation}   \label{KG}
  {(\phi _1,\phi _2)_{KG}:= {\frac{i}{2}}\int_{R^3} d^3x\, [\overline {\phi _1}
 \partial _t \phi _2\, -\partial _t \overline {\phi _1})\phi _2]}.
 \end{equation}

  Given  positive energy solutions of the free Klein-Gordon equation
  $\phi _i(x), i=1.2$, which we know to solve the quasirelativistics  equation as well, we
  realize that  ($\tilde{\phi }_i$
  is a Fourier transform of $\phi _i$)
  \begin{equation}
    (\phi _1,\phi _2)_{KG} = \int {\frac{d^3k}{k_0}}\tilde{\phi _1}^* \tilde{\phi }_2  =
        (\phi _1 , \sqrt {-\triangle +m^2}\phi _2)=    ((-\triangle +m^2)^{1/4}
  \phi _1,(-\triangle +m^2)^{1/4}
  \phi _2)
  \end{equation}
  We recall that
  $\sqrt {-\triangle +m^2}$ is a Hermitian operator in the Hilbert space  $L^2$, equipped
   with a scalar product $(\phi _1,\phi _2)= \int d^3x \overline{\phi }_1(\vec{x},t) \phi _2(\vec{x},t)$.

   We  can  now    introduce a   new  positive energy  solution for both Klein-Gordon and
  quasirelativistic Schr\"{o}dinger  equations as follows
  \begin{equation}
  \phi (x) \to  [(-\Delta +m^2)^{1/4} \phi ](x).
  \end{equation}
 At this point  we  consider  $\phi $  as the $L^2$-normalized   function in the
   domain of  $ (m^2 - \Delta )^{1/4}$, so arriving at  the {\it   mean energy } of the quantum system
   in  the state  $\phi $   (to be compared with Eq. (\ref{square})):
   \begin{equation}
  ((m^2 - \Delta )^{1/4}\phi ,(m^2 - \Delta )^{1/4}\phi )=  (\phi, (m^2 - \Delta )^{1/2}  \phi )
 = E_{\phi } >0.
   \end{equation}
 Accordingly,
    \begin{equation}
 \Phi =(1/\sqrt{E_{\phi }})  (m^2 - \Delta )^{1/4} \phi
 \end{equation}
   is    $L^2$   normalized, $(\Phi , \Phi )=1$.  (We remember that
    $(\phi , \phi )=1 \rightarrow (\phi , \phi )_{KG}= E_{\phi }$).

  The $L^2$ normalized function  $ \Phi $ may serve as a reference state  that gives account of the  energy
    spatial  distribution, c.f. \cite{barut,ibb, ibb1}. The reverse operation obviously
   reads $\phi = \sqrt{E_{\phi }} (m^2 - \Delta )^{-1/4} \Phi $.

  Introducing the pdf $\rho ({\bf x},t) = |\Phi |^2 ({\bf x},t)$ we have in hands a probability measure.
  Let $\Omega $ be a volume in $R^3$.  Then, we  interpret $p_E(\Omega ) = \int _{\Omega} d^3x \, |\Phi |^2$
   as a fraction of the total mean energy $E_{\phi }$ that is confined in the volume $\Omega$.
   Compare e.g. analogous considerations in the context of the photon wave mechanics in \cite{ibb,ibb1}.

We note in passing that the way we have introduced $\Phi $  stays in an intimate relationship
with the (induced) notion of the Newton-Wigner position operator, \cite{barut,jordan} although
this notion appears to be irrelevant for our discussion, compare also \cite{gar}.

In particular, we emphasize that the canonical quantization method adopted by us relies on the standard position-momentum pair
$[x_j,p_k] \subset \delta _{ij} I$, with $\vec{p} = -i \vec{\nabla }$.
Accordingly, most familiar  uncertainty relations hold true  $\Delta x \Delta p \geq 1/2$. One may merely enter a dispute
 of what are the states that may sharpen the Heisenberg bound.  No new position operator notions are here necessary
  (c.f.  past   discussions of the covariant position operator  notion and the causality issue in relativistic quantum theory).

Our  test model discussion  allows us to come to   a  general conclusion.  Let us disregard introductory KG equation hints (126)-(130).
Then, the mean energy-based probability measure definitions   (131), (132) can be readily   extended  (up  to dimensional coefficients)
 to  any  L\'{e}vy stable case. To this end one needs   to  begin with the mean energy definition in a given state and
  replace the quasirelativistic operator $\sqrt{m^2- \Delta }$
   in  defining formulas  (131)  and  (132)  by a  fractional operator
  $|\Delta |^{\mu /2}$ with $\mu \in (0,2)$.

   Clearly, any   connection with  standard  (relativistic)  wave equations  is lost,  with a notable exception
of the    Cauchy case $\mu =1$, where the D'Alembert link  persists.
In the latter context,   let us assume that  the  function $\Phi (k_0,{\bf k}$  in Eq. (\ref{klein})
   is selected so that under the $m \downarrow 0$ limit, $(1/|\mathbf{k}|) \Phi (|\mathbf{k}|,\mathbf{k})$ has
     the property to  vanish if we let  $|\mathbf{k}|$ go down to $0$.
 Then, all arguments
 following Eq. (\ref{klein}) retain their validity, if we replace $( - \Delta  + m^2)^{1/n}$ by  $|\Delta |^{1/2n}$
  everywhere, for  $n=1, 2$.

\subsection{Foldy's synthesis for $m>0$ and its $m=0$  extension to photon wave mechanics.}

Let us come back to the Salpeter equation in  its $m\geq 0$ versions. The link of  the $m>0$ equation  with various relatvistic equations has been
established long time ago under the name of Foldy's "synthesis of covariant particle equations", \cite{foldy}, see also \cite{simulik}.
 It has been shown that Dirac, Klein-Gordon   and Proca equations  can be reduced to a canonical form
\begin{equation}\label{form}
i \partial _t \chi = \sqrt{-\Delta + m^2}\,  \beta \, \chi
\end{equation}
where $\beta $ is a  diagonal hermitian matrix.
In particular, solutions $\phi $  of the Klein-Gordon equation  $(\Box +m^2)\phi
(\vec{x},t)=0$  yield a two-component wave function  $\chi $ with    components
 $\chi _{\pm} = (1/\sqrt{2}) [(-\Delta + m^2)^{-1/4} \partial _t\phi \mp  i(-\Delta + m^2)^{1/4}\phi ]$. The matrix $\beta $
 is identical with the Pauli matrix $\sigma _3$, comprising $1$ and $-1$ on the diagonal.

The reduction of the Dirac equation to the canonical form is accomplished by the Foldy-Wouthuysen (FW) transformation. Let $\psi $ be a solution  of
\begin{equation}
i\partial _t \psi = (\beta m  -  i\vec{\alpha }\cdot \vec{\nabla })\, \psi \, .
\end{equation}
The canonical form (\ref{form}) is obeyed by  the four-component  $\chi = U\psi $ where $U$ stands for the F-W transformation, while $\beta $ comes
from the Dirac equation  in its traditional block-diagonal form, comprising the $2\times 2$ identity matrix $I$ and its negative $-I$.

The Proca equation adopted in \cite{foldy} leads to  a canonical form with a $6$-component $\chi $ and $\beta $ being  being block-diagonal with
$3\times 3$  blocks $I$ and $-I$.

The case of mass $m=0$ has been properly addressed quite recently,  being associated with  attempts to give meaning
 to the  (single)  photon wave mechanics, \cite{ibb,ibb1,ibb2}, see also \cite{garcz}.
 The key element  there  was a reformulation of Maxwell equations  in terms of the Riemann-Silberstein vector function ${\bf F}({\bf x},t)$,
 actually  interpreted as  the  R-S wave function, e.g. a solution of the Schr\"{o}dinger-type equation with a divergence constraint
 \begin{eqnarray}
 &&i\partial _t{\bf F} = c \nabla \times {\bf F}\\
 &&\nabla \cdot {\bf F}= 0\, .
 \end{eqnarray}
The  above dynamical equation can be given   more explicit Schr\"{o}dinger form  by invoking spin $1$  ($3\times 3$, no $\hbar $ involved)
 matrices   $[S_j,S_k] = i\epsilon _{jkl} S_l$, so that c.f. \cite{ibb}
 \begin{equation}
i\partial _t{\bf F} = {\frac{c}{i}}  ({\bf S}\cdot  \nabla )  {\bf F}
 \end{equation}
We note that the divergence condition excludes  the potentially troublesome   $\vec{k}=\vec{0} $ mode.  Therefore we can safely introduce
 the inverse of the Cauchy operator  $|\Delta |^{1/2}$, while acting upon $\bf F$  such that $\nabla \cdot {\bf F}= 0$.

By introducing the helicity operator  $\hat{\Lambda }$  for photons (loosely speaking, a projection of spin on momentum) which
is a manifestly nonlocal operator, we can rewrite  the photon Hamiltonian $\hat{H} = \hat{\bf p}\cdot \hat{\bf S}$ as follows
\begin{eqnarray}
&&\hat{H} = c |\hat{\bf p}|\,  \hat{\Lambda }\\
&& \hat{\Lambda }= {\frac{1}{|\nabla |}}
 (\hat{\bf p}\cdot \hat{\bf S})
\end{eqnarray}
where  $\hat{\bf p} = - i \nabla $ and  we interpret $|\hat{\bf p}|$ as  $|\nabla |= |\pm \Delta |^{1/2}$.  We can pass to the helicity basis (with
the corresponding eigenvalues $\pm 1$), ending with positive energy solutions of the $m=0$ variant of the Salpeter equation, in the canonical form
\begin{equation}
i\partial \chi  = c |\Delta |^{1/2}\beta  \chi
\end{equation}
where $\beta $ is the  $6\times 6$ diagonal matrix, encountered before in connection with the Proca equation. Here,   $\chi $ is a $6$-component vector
composed of  positive frequency pieces taken form the original  R-S vector  and its complex adjoint, i. e. ${\bf F}^{(+)}$ and ${\bf F}^{(-)*}$   respectively.
Compare a discussion of the helicity /spin  spectral issues  in Section 2.1 of \cite{ibb}, see also \cite{garcz}.

In the helicity basis,  the photon wave function dynamics is   {\it   nonlocal} and shows  all intriguing features  (radial  expansion) discussed before
in subsection III.F.1.  We note that the  expanding/contracting  behavior    has been
previously associated with  the photon energy density  ${\bf F}\cdot {\bf F}^*$, see e.g. Fig. 3 in \cite{ibb1}.
 A discussion of  involved  uncertainty relations, from varied perspectives, including the mean  energy normalization of the wave function,
  can be found in \cite{ibb,ibb1,ibb2}, see specifically   Eqs. (5.28)-(5.31) of Ref. \cite{ibb}).  Their obvious  validity is not that
   illuminating as a useful  indicator  delocalization in view of   rather rapid    radial  expansion of wave packets.

 By invoking  the scalar product considerations of Section 5.1 in \cite{ibb} we  readily arrive at the $m=0$ version of the interplay  between solutions of the second order wave equations  (here, d'Alembert) and those inferred from the Schr\"{o}dinger type equation.
Namely, the inversion relation $\phi = \sqrt{E_{\phi }} (m^2 - \Delta )^{-1/4} \Phi $  of subsection IV.B has a counterpart in the photon
wave mechanics. Up to  the  mean energy normalization, the
Landau-Peierls wave function  $\phi = |- \Delta |^{-1/4} \Phi $   (c.f. section 5.3 in \cite{ibb})  is  precisely the  outcome of a mapping
of  solutions of the d'Alembert equation, $({\frac{1}{c^2}}{\frac{\partial ^2}{\partial t^2}}  - \Delta ) {\bf F}_{\pm} =0$,
 (\ref{klein})  (set m=0 therein),   to  $L^2 $ normalized solutions of
 Eqs.  (138) and /or  (141).

 To be more explicit, we note that all previous steps (129) through (130) can be safely  repeated if:   (i) in view of $m=0$, we  replace
  $(-\triangle +m^2)^{1/2n}$ by $|\triangle |^{1/2n}$, with $n=1,2 $, (ii)  next  extend the  interpretation of  integration   to
  encompass summmation over helicity indices  $\lambda = 1, -1$,  and (iii) insert $\Phi _{\lambda }(k_0,\vec{k})$ instead of the previous
$\Phi (k_0,\vec{k})$.

Then  the ($m\downarrow 0$) KG scalar product (\ref{KG})   assumes  the d"Alembertian form  and refers to two-component
 positive energy solutons of both d'Alembert  and  the Cauchy-Schr\"{o}dinger  equation (here we consider the squared  norm of $\phi $ ):
\begin{equation}
(\phi ,\phi )_{d'A} =  \sum _{\lambda } \int {\frac{d^3k}{k_0}}\tilde{\phi _{\lambda }}^* \tilde{\phi }_{\lambda }  =
      (\phi  , |\triangle |^{1/2}\phi )=    (|\triangle |^{1/4} \phi ,|\triangle |^{1/4} \phi )= E_{\phi }
\end{equation}
and the $L^2$ normalization of  a  positive energy solution   $(\phi , \phi )=1$
clearly implies  $(\phi , \phi )_{d'A}=E_{\phi } $   and the $L^2$  normalization  $({\bf \Phi },{\bf  \Phi })=1$ of
\begin{equation}
{\bf \Phi }=(1/\sqrt{E_{\phi }})  | \Delta |^{1/4} \phi   \, .
\end{equation}

Now we are exactly at the point where the Landau-Peierls  (LP) wave function happens to be introduced.  This issue was a subject of disagreements
in the literature, that ranged form accepting to discard th L-P functions as an object of a physical relevance, \cite{good,ibb}.
 On the basis of  (142), (143),  we know that both ${\bf \Phi }$ and $\phi $ are  fairly legitimate objects.

  However  if   ${ \bf \Phi }$   is what   we interpret    to provide  a consistent probabilistic interpretation  of a
  single photon quantum state (see e.g. also sections 5 and 5 of \cite{ibb}  and \cite{good}), then  we may always reproduce $\phi $ with the squared  norm
 $(\phi ,\phi )_{d'A} =  \sum _{\lambda } \int {\frac{d^3k}{k_0}}\tilde{\phi _{\lambda }}^* \tilde{\phi }_{\lambda }$  according to
   $\phi = \sqrt {E_{\phi }} |\Delta |^{-1/4} \Phi $.
   We recall that   $(\phi ,\phi ) =1 \rightarrow (\phi , \phi )_{d'A} = E_{\phi }$. Hence $\phi / \sqrt{E_{\phi }}$  is d'Alembert-normalized.
   That can also  be accomplished by writing
   \begin{equation}
   {\frac{1}{E_{\phi }}} (\phi , \phi )_{d'A} =  (|\Delta |^{-1/4} \phi,  |\Delta |^{-1/4} \phi )_{d'A} = 1    \, .
   \end{equation}

  We note that  often expressed   in the literature  past  prejudices, concerning the  LP wave function nonlocality
   (reaching back to W. Pauli, (1933), \cite{ibb}),  in the light of our discussion must be taken under scrutiny as
   invalid objections.   All ingredients of our  theory of  quantum motion  are {\it spatially  nonlocal}.
   That is  possibly  somewhat  blurred  on the Fourier level  of description, but  fairly  obvious  while invoking
    the primary level of of the theory, based on the exploitation of    non-Gaussian noise (L\'{e}vy)   generators.

\section{Outlook}

    The main message of the present paper is a comprehensive analysis of the intimate relationship between jump-type
    stochastic processes (like e.g. L\'{e}vy flights)   and nonlocal (due to integro-differential operators involved)
    quantum dynamics.
  In the course  of the paper we were  very explicit in deriving certain formulas (like e.g. those for  various propagators) which
   are   often   presented in the faulty
   form in the current literature.  We have  also attempted to clarify a number of topics related to so-called fractional quantum mechanics,
   which as well is  not free  of drawbacks and plainly invalid  formulas, \cite{luchko,kwasnicki}.
   Such situaton is  in part   a consequence of   serious  technical  difficulties. Those in fact  have prohibited the pertinent theory
     from   gaining  popularity in the research literature  of the 70ties, c.f. \cite{gar}
     for earlier references.

 Another  serious problem  pertains to  quite appealing  and  both  customarily   and uncritically  adopted
    Fourier representation of   the   spatially nonlocal  quantum dynamics. This  is convincingly presented
     in the literature  on various aspects of relativistic wave mechanics and that of quantum fields, \cite{drell,drell1}),
     where  the Fourier analysis  has been always   regarded  as a legitimate  departure point in the development of
      quantum theory.

In the present paper   the Fourier representation   of  analyzed  dynamical   problems     is never  a primary,
   but a secondary,  derived  notion.  It does not  provide a fully
    adequate   transcription of   the   original   spatially nonlocal formalism, unless employed with due care.

  Our  input of the theory of random motion, specifically  that   of jump-type  heavy-tailed probability laws,
     so far  has not received the status of  relevance   in the    quantum context. Presumably it was the reason of why the inherent
  nonlocality of  the quasirelativistic  dynamics  (take that as an exemplary case) has not been  properly   addressed.
       We   know about one  isolated  attempt  in  this direction   due to  \cite{lammerzahl}).

As discussed  in detail, the nonlocal quantum dynamics  may lead to  strongly   enhanced   spatial   delocalization
 of any "narrow-peaked"  initial probability density function. This is not  merely a familiar wave packet spreading of the
 standard  Schr\"{o}dinger quantum dynamics.   The  pertinent  delocalization   happens to be  conspicuously    amplified, like e.g.
  in  case of   simple  relativistic  physics-motivated  models. That is  exemplified  by
  a dynamically generated   multi-modality emerging    out   of  initially unimodal   pdfs  in the 1D  case.
   In the 3D case the  quantum  dynamics     induces a  radial expansion  of
  any initially localized  pdf.   The speed of that expansion
   from the vicinity of the origin, in the massless case  equals the velocity of light   $c$.
    This phenomenon makes doubtful a diagnostic utility of  standard   uncertainty   relations.

   A point  worth mentioning   is that of a predominantly   lacking "particle" concept   of the
  nonlocal  quantum   dynamics.  Doubts pertaining to "what is actually jumping there"   have been expressed
  before \cite{gar} and refer  to the massive   Salpeter case  as well, see also \cite{angelis}.
   Interestingly, the  same "what is jumping" problem persists  in the purely classical framework of L\'{e}vy stochastic processes.

 From the jump-processes perspective  (which are nonlocal  phenomena as well) one  should be
  be very careful while  interpreting  the associated    transport  phenomena  in terms of "particles".
 All L\'{e}vy stable-supported dynamical scenarios  are devoid of any  genuine "massive  particle"
  connotations.  Presently, our view is that it is not a "particle"
 that  undergoes a jump-type process, although other opinions have been expressed in the past, \cite{gar,angelis}.

 Since the Salpeter  ($m\geq0$)   case   belongs to our non-Gaussian (and non-Laplacian) inventory,
 the  very  "particle"   position  (operator)
 concept  needs to  be  put  under scrutiny, including past vigorous discussions of the causality problem in relativistic quantum mechanics and  that of
  of a  covariant position operator,   or position operator at all in the photon context.
    The "photon position"  and related localization   concept,  appears to be useless in the developed framework
    and turns out  to stay in  a deep  contradiction with the radial expansion of
    photon wave packets (see below).

    Interestingly,   our methods   find  some support in investigations of the photon wave function
     (and  photon  wave  mechanics). Specifically while combining  the time evolution  of   wave packets with
  various  forms of uncertainty relations proposed so far, \cite{ibb,ibb1, ibb2}, see also \cite{wiese}.

\end{document}